\documentclass[structabstract]{aa}  
\usepackage{graphicx}
\usepackage{lscape}
\usepackage{amsmath}
\usepackage{txfonts}
\usepackage{url}
\usepackage{natbib}
\usepackage[usenames,dvipsnames,svgnames,table]{xcolor}

\begin{document}
   \title{Grids of stellar models with rotation}

   \subtitle{III. Models from 0.8 to 120 $M_{\sun}$ at a metallicity \textit{Z} = 0.002}
   
   \titlerunning{Grids of stellar models with rotation. III}

   \author{C. Georgy\inst{\ref{inst1},\ref{inst2}} \and S. Ekstr\"om\inst{\ref{inst3}} \and P. Eggenberger\inst{\ref{inst3}} \and G. Meynet\inst{\ref{inst3}}  \and L. Haemmerl\'e\inst{\ref{inst3}} \and A. Maeder\inst{\ref{inst3}} \and A. Granada\inst{\ref{inst3}} \and J. H. Groh\inst{\ref{inst3}} \and R.~Hirschi\inst{\ref{inst1},\ref{inst4}} \and N. Mowlavi\inst{\ref{inst3}} \and N. Yusof\inst{\ref{inst6},\ref{inst7}} \and C. Charbonnel\inst{\ref{inst3},\ref{inst5}} \and T. Decressin\inst{\ref{inst3}} \and F. Barblan\inst{\ref{inst3}}}

   \authorrunning{Georgy et al.}

   \institute{Astrophysics group, EPSAM, Keele University, Lennard-Jones Labs, Keele, ST5 5BG, UK\\
                   \email{c.georgy@keele.ac.uk}\label{inst1} \and
                   Centre de recherche astrophysique, Ecole Normale Sup\'erieure de Lyon, 46, all\'ee d'Italie, F-69384 Lyon cedex 07, France\label{inst2} \and
                   Geneva Observatory, University of Geneva, Maillettes 51, CH-1290 Sauverny, Switzerland\label{inst3} \and
                   Institute for the Physics and Mathematics of the Universe (WPI), University of Tokyo, 5-1-5 Kashiwanoha, Kashiwa, 277-8583, Japan\label{inst4} \and
                   IRAP, UMR 5277 CNRS and Universit\'e de Toulouse, 14, Av. E.Belin, 31400 Toulouse, France\label{inst5} \and
                   Department of Physics, Faculty of Science, University of Malaya, 50603 Kuala Lumpur, Malaysia\label{inst6} \and
                   Quantum Science Center, Faculty of Science, University of Malaya, 50603 Kuala Lumpur, Malaysia\label{inst7}}
                   
   \date{Received ; accepted }

 \abstract
   {} 
   {We study the impact of a subsolar metallicity on various properties of non-rotating and rotating stars, such as surface velocities and abundances, lifetimes, evolutionary tracks, and evolutionary scenarios.}
   {We provide a grid of single star models covering a mass range of 0.8 to 120 $M_{\sun}$ with an initial metallicity $Z=0.002$ with and without rotation. We discuss the impact of a change in the metallicity by comparing the current tracks with models computed with exactly the same physical ingredients but with a metallicity $Z=0.014$ (solar).}
   {We show that the width of the main-sequence (MS) band in the upper part of the Hertzsprung-Russell diagram (HRD), for luminosity above $\log\left(L/L_{\sun}\right) > 5.5$, is very sensitive to rotational mixing. Strong mixing significantly reduces the MS width. Here for the first time over the whole mass range, we confirm that surface enrichments are stronger at low metallicity provided that comparisons are made for equivalent initial mass, rotation, and evolutionary stage. We show that the enhancement factor due to a lowering of the metallicity (all other factors kept constant) increases when the initial mass decreases. Present models predict an upper luminosity for the red supergiants (RSG) of $\log\left(L/L_{\sun}\right)$ around $5.5$ at $Z=0.002$ in agreement with the observed upper limit of RSG in the Small Magellanic Cloud. We show that models using shear diffusion coefficient, which is calibrated to reproduce the surface enrichments observed for MS B-type stars at $Z=0.014$, can also reproduce the stronger enrichments observed at low metallicity. In the framework of the present models, we discuss the factors governing the timescale of the first crossing of the Hertzsprung gap after the MS phase. We show that any process favouring a deep localisation of the H-burning shell (steep gradient at the border of the H-burning convective core, low CNO content), and/or the low opacity of the H-rich envelope favour a blue position in the HRD for the whole, or at least a significant fraction, of the core He-burning phase.}
   {}
 

\maketitle

\defcitealias{Ekstrom2012a}{Paper I}
\defcitealias{Maeder2001a}{MM01}
\defcitealias{Maeder1997a}{M97}
\defcitealias{Talon1997a}{TZ97}
\section{Introduction}

Many physical properties of stars depend on the initial metallicity, such as the mass loss by stellar winds, the equation of state, the opacity, and the nuclear energy generation rate. In addition to these \textit{classical} dependencies, some others may be recognised in the future. For instance, the initial angular momentum content of stars on the zero-age main-sequence (ZAMS), or the proportion of stars in close binary systems may somehow depend on the metallicity. Concerning the impact of rotation on the evolution of low-metallicity massive stars, several effects have already been found in previous works \citep[\textit{e.g.}][]{Maeder2001a,Meynet2002a,Brott2011a,Yusof2013a}:
\begin{itemize}
\item At low metallicity, stars are more compact, the gradients of the angular velocity are steeper, and thus the mixing of the chemical elements is more efficient.
\item Since the mass-loss rates by radiatively line-driven winds are weaker at low metallicity, less angular momentum is lost by this process at the surface, and therefore stars keep a larger amount of their initial angular momentum content in their interiors. The most massive stars thus show higher surface velocities than do their more metal-rich siblings, which seems to be supported by observations \citep{Martayan2007a}.
\end{itemize}

The effects mentioned above have been studied for stars more massive than about $9\,M_{\sun}$. In the present work, we explore to what extent similar consequences are also attained in the low and intermediate mass range \citep[see also][]{Georgy2013a}. Although we focus here on single star models, the physics of rotation is also relevant for modelling stars in close binaries. Indeed, any close binary evolution model needs a well constrained theory for the effects of rotation. Single or non-interacting stars (which may be in binary systems) are certainly more appropriate to checking this physics than interacting binaries, where these effects are mixed with other ones.

This grid is part of a larger database of stellar models. At the moment this database contains models for the metallicity $Z=0.014$ \citep[solar metallicity,][hereafter Paper I]{Ekstrom2012a} computed with exactly the same physics. This grid thus offers an ideal comparison basis to discuss the impact of rotation at a lower metallicity.

The paper is structured as follows. The physical ingredients of the models are discussed in Sect.~\ref{SecPhymod}. The computed stellar models and the electronic tables, which we make publicly available via a web page, are presented in  Sect.~\ref{SecTables}. The properties of the non-rotating (or initially slowly rotating) tracks are discussed in Sect.~\ref{SecResultsNOROT}, while the rotating models are presented in Sect.~\ref{SecResultsROT}. We briefly conclude in Sect.~\ref{SecDiscu}.

\section{Physical ingredients of the models \label{SecPhymod}}

The physical ingredients used in this paper are exactly the same as in \citetalias{Ekstrom2012a}, and are briefly summarised below. For the few cases where the change in the metallicity induced changes in the physics of our models, it is explicitly mentioned in the text.

The initial abundances of H, He, and metals are set to $X=0.747$, $Y=0.251$, and $Z=0.002$ \citep[see][for more details on how these quantities have been chosen]{Georgy2013a}. The mixture of heavy elements is assumed to be the same as in \citetalias{Ekstrom2012a}, the absolute abundances being just scaled to the metallicity considered here. This choice makes direct comparisons with the observed surface abundances of stars in the Small Magellanic Cloud (SMC) a little tricky because in the SMC, the N/C and N/O ratios measured in HII regions (considered to be equivalent to the initial abundances of stars recently formed in the SMC) are much lower than in the Sun \citep{Heap2006a,Hunter2009a}. We shall come back to this point in Sect.~\ref{SubSecSurfAbund}.  For all other purposes (age determinations, studies of stellar populations in  the SMC to mention just a few), the present tracks can be used without any restrictions. 

The opacities are generated with the OPAL tool\footnote{\url{http://adg.llnl.gov/Research/OPAL/opal.html}} \citep[based on][]{Iglesias1996a} for this particular mixture. They are complemented at low temperatures by the opacities from \citet{Ferguson2005a} adapted for the high Ne abundance used in this work \citep{Cunha2006a}.

Solar-type models with $M < 1.25\,M_{\sun}$ are computed with the OPAL equation of state \citep[EOS,][]{Rogers2002a}. For the higher mass models, the EOS is that of a mixture of perfect gas and radiation, and accounts for partial ionisation in the outermost layers, as in \citet{Schaller1992a}, and for the partial degeneracy in the interior in the advanced stages.

The nuclear reaction rates are the same as in \citetalias{Ekstrom2012a}. They are taken mainly from the Nacre database \citep{Angulo1999a}, although some have been redetermined more recently and updated \citepalias[see details in][]{Ekstrom2012a}. The models of massive stars ($M > 9\,M_{\sun}$) were computed by incorporating the NeNa-MgAl cycle. The energy loss in plasma, pair, and photo-neutrinos processes are taken from \citet{Itoh1989a,Itoh1996a}.

The convective zones are determined with the Schwarzschild criterion. For the H- and He-burning phases, the radius of the convective core is extended with an overshoot parameter $d_\text{over}=0.1H_P$ for $1.7\,M_{\sun}$ and above, $0.05H_P$ between $1.25$ and $1.5\,M_{\sun}$, and $0$ below, where $H_P$ is the pressure scale height $-\frac{\text{d}r}{\text{d}P}P$. To avoid an extension of the core greater than $10\%$, the extension of the radius is limited to $R_\text{cc}(1+d_\text{over}/H_P)$ in case $H_P$ is greater than the radius of the convective core $R_\text{cc}$ (without overshooting). The value of the overshoot parameter was calibrated in the mass domain $1.35-9\, M_{\sun}$ at solar metallicity to ensure that the rotating models closely reproduce the observed width of the MS band. In convective zones, due to the very short turn-over time compared to the nuclear timescale, the chemical mixing is assumed to be instantaneous. Moreover, the convective zones are supposed to rotate as a solid body.

In the low-mass stars range ($M\leq1.25 M_{\sun}$), the outer convective zone is treated according to the mixing length theory, with a solar calibrated value for the mixing-length parameter \citep[$\alpha_\text{MLT} \equiv \ell/H_P=1.6467$, $\ell$ being the mixing length, see \textit{e.g.}][]{Kippenhahn1990a}. For stars with $M>1.25\, M_{\sun}$, the difference in the EOS implies a slightly lower value for this parameter: $\alpha_\text{MLT}=1.6$. 

For the most luminous models, the turbulence pressure and acoustic flux need to be included in the treatment of the envelope. As in \citet{Schaller1992a}, this is done according to \citet[see also \citealt{Maeder2009a}, Sect.~5.5]{Maeder1987c}, using a mixing length taken on the density scale: $\alpha_\text{MLT}=\ell/H_\rho = \ell (\alpha - \delta \nabla) / H_P = 1$. The use of $H_\rho$ instead of $H_P$ removes the risk of having an unphysical density inversion in the envelope \citep{Stothers1973a}. The side effect of this treatment is that the redwards extension of the tracks in the Hertzsprung-Russell diagram (HRD) is reduced by $0.1-0.2\,\text{dex}$ in $T_\text{eff}$ \citep[see][]{Maeder1987b}. For this reason, we restrict the use of $H_\rho$ only to the models with $M\geq40\,M_{\sun}$, which naturally do not extend to the extreme red part of the HR diagram.

Atomic diffusion due to concentration and thermal gradients is included for low-mass models exhibiting an effective temperature lower than a given value on the ZAMS. This limit is fixed to an effective temperature on the ZAMS of $6300\,\text{K}$, which corresponds to the limit of $1.25\, M_{\sun}$ used for the computation of the grid at solar metallicity \citepalias{Ekstrom2012a} and to $1\, M_{\sun}$ in this work. However, the radiative acceleration is neglected in our computations. This is well justified for low-mass stellar models with extended convective envelopes \citep{Turcotte1998a}. For stars with shallow convective envelopes, when only atomic diffusion due to concentration and thermal gradients is included in the computation, helium and heavy elements are rapidly drained out of the envelope leading to surface abundances in obvious contradiction with spectroscopic observations. This indicates that another transport mechanism needs to be introduced for these models (radiative acceleration, rotational mixing, \textit{etc.}) to counteract the effects of diffusion due to concentration and thermal gradients in the external stellar layers \citep[see for instance Fig.~10 of][]{Carrier2005a}. In the present models, the diffusion due to concentration and thermal gradients is thus simply not included for these stars and models more massive than a given limit are then computed without atomic diffusion. This is also motivated by recalling that for more massive stars the typical timescale associated to atomic diffusion becomes rapidly much longer than the main-sequence (MS) lifetime.

Magnetic braking of the stellar surface is taken into account for models having an effective temperature on the ZAMS lower than $7900\,\text{K}$, which corresponds to the limit of $1.7\,M_{\sun}$ used at solar metallicity \citepalias{Ekstrom2012a} and to $1.35\, M_{\sun}$ in this work. This braking is taken into account using a solar calibrated value for the braking constant.

The treatment of rotation is the same as in \citetalias{Ekstrom2012a} and accounts for the transport of the angular momentum and chemical species by the meridional currents and shear turbulence. Models are assumed to be in a shellular rotation state, \textit{i.e.} each shell rotates uniformly at all latitudes. For the rotating models, we started with a solid-body rotation on the ZAMS. The justification of this choice is that meridional circulation will adjust the rotational profile to an equilibrium profile \citep{Zahn1992a,Denissenkov1999a,Meynet2000a} in a short timescale compared to stellar evolution, so that we expect the effects of the previous evolution to be rapidly cancelled. Interestingly, we show in Sect.~\ref{SubSecVelocities} that indeed the rotational profile on the ZAMS resulting from a fully consistent computation of the pre-main-sequence phase (PMS) with accretion is very near from a flat rotational profile, providing thus an additional justification for starting with a solid-body rotation. Advection is followed during the MS. In later stages, because of the shortest evolution time scale, it has no time to efficiently transport angular momentum, and we thus account for it in its diffusion approximation, which gains in computational simplicity and efficiency. Our models do not take the effects of transport due to internal magnetic field into account.

Similarly, the mass-loss rates are implemented as in \citetalias{Ekstrom2012a}\footnote{\footnotesize{Note however that in contrast with the $Z_{\sun}$ case, the increased mass-loss rates we use during the RSG phase for stars more massive than $20\,M_{\sun}$ has only a negligible effect in this work due to the very short duration of the RSG phase (see below).}}. The only difference is that we accounted for a metallicity dependence of the mass loss rates with the initial metallicity according to the following rules:
\begin{itemize}
\item during the MS phase and blue supergiant phase, for stars more massive than $7\,M_{\sun}$, we use a scaling of the mass-loss rates with the metallicity of the form $\dot M(Z) =(Z/Z_{\sun})^\alpha \dot M(Z_{\sun})$. The parameter $\alpha$ is set to $0.85$ when the \citet{Vink2001a} mass-loss rate is used, and to $0.5$ when the \citet{deJager1988a} mass-loss rate is used;
\item for stars with effective temperatures lower than $3.7$ (red giants and supergiants) no metallicity dependence of the mass-loss rates is applied. Indeed, according to \citet{vanLoon2005b} and \citet{Groenewegen2012a,Groenewegen2012b} the metallicity dependence of the mass-loss rates for these stars do appear weak at best;
\item during the Wolf-Rayet (WR) phase, we follow the metallicity dependence proposed by \citet{Eldridge2006a}, with $\dot M(Z) =(Z/Z_{\sun})^{0.66} \dot M(Z_{\sun})$.
\end{itemize}

\section{The stellar models and electronic tables \label{SecTables}}

We present models of $0.8$, $0.9$, $1$, $1.1$, $1.25$, $1.35$, $1.5$, $1.7$, $2$, $2.5$, $3$, $4$, $5$, $7$, $9$, $12$, $15$, $20$, $25$, $32$, $40$, $60$, $85$, and $120\,M_{\sun}$. For each mass, we computed both a rotating and a non-rotating model.

The rotating models start on the ZAMS with a value of $V_\text{ini}/V_\text{crit}=0.4$\footnote{\footnotesize{The critical velocity is defined as $V_\text{crit} = \left(\frac{2GM}{3R_\text{pb}}\right)^\frac{1}{2}$ with $R_\text{pb}$ the polar radius at the break-up velocity \citep[see][]{Maeder2000a}.}}. This choice is based on the peak of the velocity distribution of young solar-metallicity B stars in \citet[see their Fig.~6]{Huang2010a}. As already recalled above \citep[see also][]{Maeder2001a}, at low metallicity, stars are more compact and have a higher critical velocity at a given mass. Therefore a given value of $V_\text{ini}/V_\text{crit}$ on the ZAMS corresponds to a higher rotational velocity during the MS phase. This initial rotation rate corresponds to a mean MS velocity between $150$ and $320\,\text{km}\cdot\text{s}^{-1}$ for the stars that are not magnetically braked ($M_\text{ini} \ge 1.7\,M_{\sun}$). The corresponding velocity obtained in our previous grid at solar metallicity lays between $110$ and $220\,\text{km}\cdot\text{s}^{-1}$.

In the high  mass range, this velocity is higher than the peak of the observed velocity distribution of MS stars in the SMC, which is around $150-180\,\text{km}\cdot\text{s}^{-1}$ for unevolved  O-type stars \citep{Mokiem2006a}. Thus present models may be more representative of the stars at the high end of the  velocity distribution rather than of the average rotators. Further computations will be performed to complete the present grid in order to well cover the observed average rotation rates. However present models are interesting to study the impact of low metallicity at a given $V_\text{ini}/V_\text{crit}$, as they allow for a direct comparison with the solar metallicity grids \citepalias{Ekstrom2012a}.

The models are evolved up to the end of the core carbon burning ($M_\text{ini} \ge 12\,M_{\sun}$), the early asymptotic giant branch ($2.5\,M_{\sun} \le M_\text{ini} \le 9\,M_{\sun}$), or the helium flash ($M_\text{ini} \le 2\,M_{\sun}$).

Electronic tables of the evolutionary sequences are available on the web\footnote{See the webpage\\\url{http://obswww.unige.ch/Recherche/evol/-Database-}\\or the CDS database at\\\url{http://vizier.u-strasbg.fr/viz-bin/VizieR-2}.}. For each model, the evolutionary track is described by 400 selected data points, each one corresponding to a given stage. Each point of different evolutionary tracks with the same number corresponds to similar stages to facilitate the interpolation between the tracks. The points are numbered as described in \citetalias{Ekstrom2012a}.

We also developed an interactive web application, giving the possibility to compute and download evolutionary tracks for any given choice of the initial mass, metallicity and rotation, these values being  framed by the values of the present stellar model grids and those of  \citetalias{Ekstrom2012a}, \citet{Georgy2013a} and \citet{Mowlavi2012a} . This application can also provide isochrones for any age for a given metallicity and initial rotation rate. It can be found at \url{http://obswww.unige.ch/Recherche/evoldb/index/}.

Table \ref{TabListModels} presents the general characteristics of all the models at the end of each burning phase. After the initial mass, initial velocity, and mean velocity on the MS (col. 1 to 3), we give for each burning phase its duration (col. 4, 10, and 16), the actual mass of the model (col. 5, 11, and 17), the equatorial velocity (col. 6, 12, and 18), the surface He abundance in mass fraction (col. 7, 13, and 19), and the surface abundances ratios N/C (col. 8, 14, and 20) and N/O (col. 9, 15, and 21), both in mass fraction. When calculating the lifetimes in the central burning stages, we consider the start of the stage as the time when 0.003 in mass fraction of the main burning fuel is burnt. We consider that a burning stage is finished when the main fuel mass fraction drops below $10^{-5}$. 

\section{Properties of the non-rotating models \label{SecResultsNOROT}}

\begin{figure*}
\centering
\includegraphics[width=.9\textwidth]{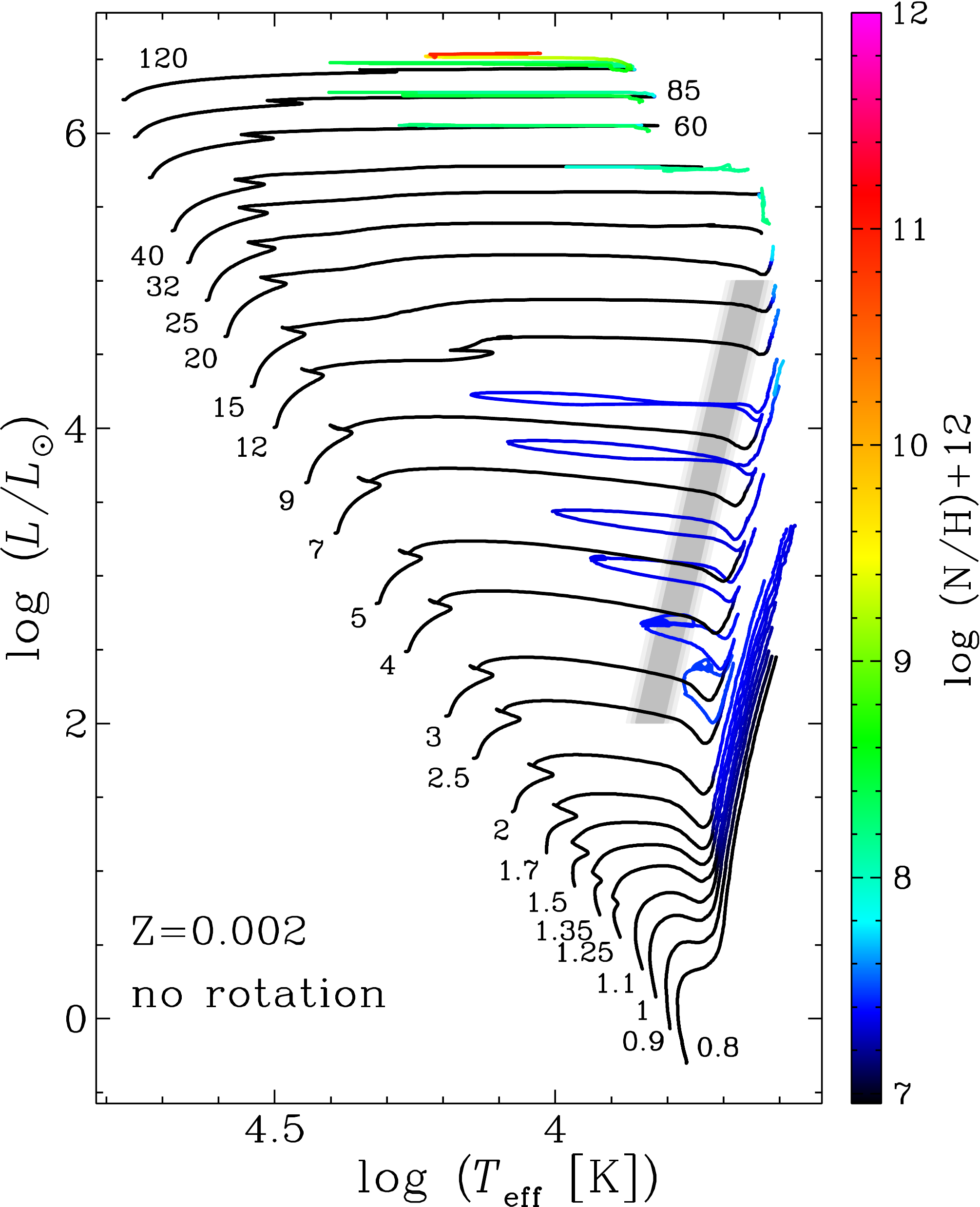}
\caption{HRD for the non-rotating models. The colour scale indicates the surface number abundance of nitrogen on a logarithmic scale where the abundance of hydrogen is $12$. The grey shaded area represents the Cepheid instability strip for the SMC according to \citet{Tammann2003a}.}
\label{Fig_HRDnorot}
\end{figure*}

In this Section, we compare the present non-rotating tracks at $Z=0.002$ shown in Fig.~\ref{Fig_HRDnorot} with the corresponding non-rotating models computed at $Z=0.014$ presented in \citetalias{Ekstrom2012a}. Many of the effects described below have already been discussed in previous works \citep[see \textit{e.g.} the review by] []{Chiosi1986a,Bertelli2008a}. We note the following differences  between these two sets of models (going from the upper to the lower part of the HRD):
\begin{itemize}
\item In the present grid,  the MS band widens in the upper part of the HRD and reaches its greatest extension for the highest mass of our sample, the $120\,M_{\sun}$. At solar metallicity, the MS band widens between $40$ and $120\,M_{\sun}$, with a maximal extension at around $60\,M_{\sun}$. The widening is linked to the efficiency of stellar winds. If the winds are strong and the mass lost during the MS is significant, the mass fraction of the core increases with respect to the total mass (like an overshoot would do). However if the mass loss is still stronger and sufficient to remove a significant part of the H-rich envelope and keep the star in a blue position in the HRD, the MS width shrinks back. These effects produce the elbow appearing in different mass ranges depending on the strength of the stellar winds. At $Z_{\sun}$, the shrinking occurs for stars with $M_\text{ini}>60\,M_{\sun}$. At low $Z$, mass-loss rates are globally weaker, hence only the widening occurs in the highest masses range.
\item The non-rotating tracks at $Z=0.002$ give a maximum luminosity for the red supergiants (RSGs) around $5.7$, the same value as the $Z=0.014$ non-rotating models. However, the value of $5.7$ at $Z=0.002$ is obtained by the $40\,M_{\sun}$ model that spend only a very short time as a RSG (about $1000\,\text{yr}$). Thus the effective upper luminosity is rather in the range of $5.5$ corresponding to the luminosity reached by the $32\,M_{\sun}$ stellar model during its RSG phase (which lasts $45000\,\text{yr}$).
\item Models with $M_\text{ini}\geq40\,M_{\sun}$ at $Z=0.002$ evolve back to the blue after a luminous blue variable (LBV) or a RSG phase. The corresponding inferior mass limit for evolving back to the blue is $25\,M_{\sun}$ for the non-rotating solar metallicity models. As a consequence of weaker stellar winds, this mass limit is increased at low $Z$. We shall discuss this point in more details in Sect.~\ref{SubSecHRDROT}.
\item Extended blue loops, crossing the Cepheid instability strip\footnote{we take the limits of the instability strip given by \citet{Tammann2003a}}, occur for masses between $3$ and $9\,M_{\sun}$ at $Z=0.002$. At solar metallicity, the mass range presenting such a blue loop lies between $5$ and $9\,M_{\sun}$. The corresponding luminosity ranges are $2.5\lesssim \log(L/L_{\sun})\lesssim4.25$ at $Z=0.002$ and $2.75\lesssim \log(L/L_{\sun})\lesssim4.25$ at $Z_{\sun}$.
\item The tip of the red giant branch at $Z=0.002$, \textit{i.e.} the luminosity at which the He-flash occurs is around $\log(L/L_{\sun}) = 3.5$ for stars in the mass range between $1.35$ and $1.7\,M_{\sun}$, slightly more luminous than at solar metallicity (around $3.4$). 
\item As already found in previous non-rotating models \citep[\textit{e.g.}][]{Schaller1992a}, central He-burning at low metallicity occurs mostly in the blue part of the HRD, in contrast with what happens at solar metallicity.
\end{itemize}

In Figure \ref{Fig_HRDnorot}, the colour code indicates the ratio of the nitrogen to the hydrogen number abundance at the surface of the stars, on a logarithmic scale (the initial value is $\log(\text{N/H})+12 = 6.95$). No change in the surface abundances is expected during the MS phase for all the stars considered here. For masses below about $40\,M_{\sun}$, the changes in the surface abundances occur after the star has passed through the RSG  or red giant stage (first dredge-up). At solar metallicity, stars with $M_\text{ini}>60\,M_{\sun}$ show some nitrogen enhancements already at the end of the MS phase because of the strong mass loss.

\begin{figure*}
\centering
\includegraphics[width=.98\textwidth]{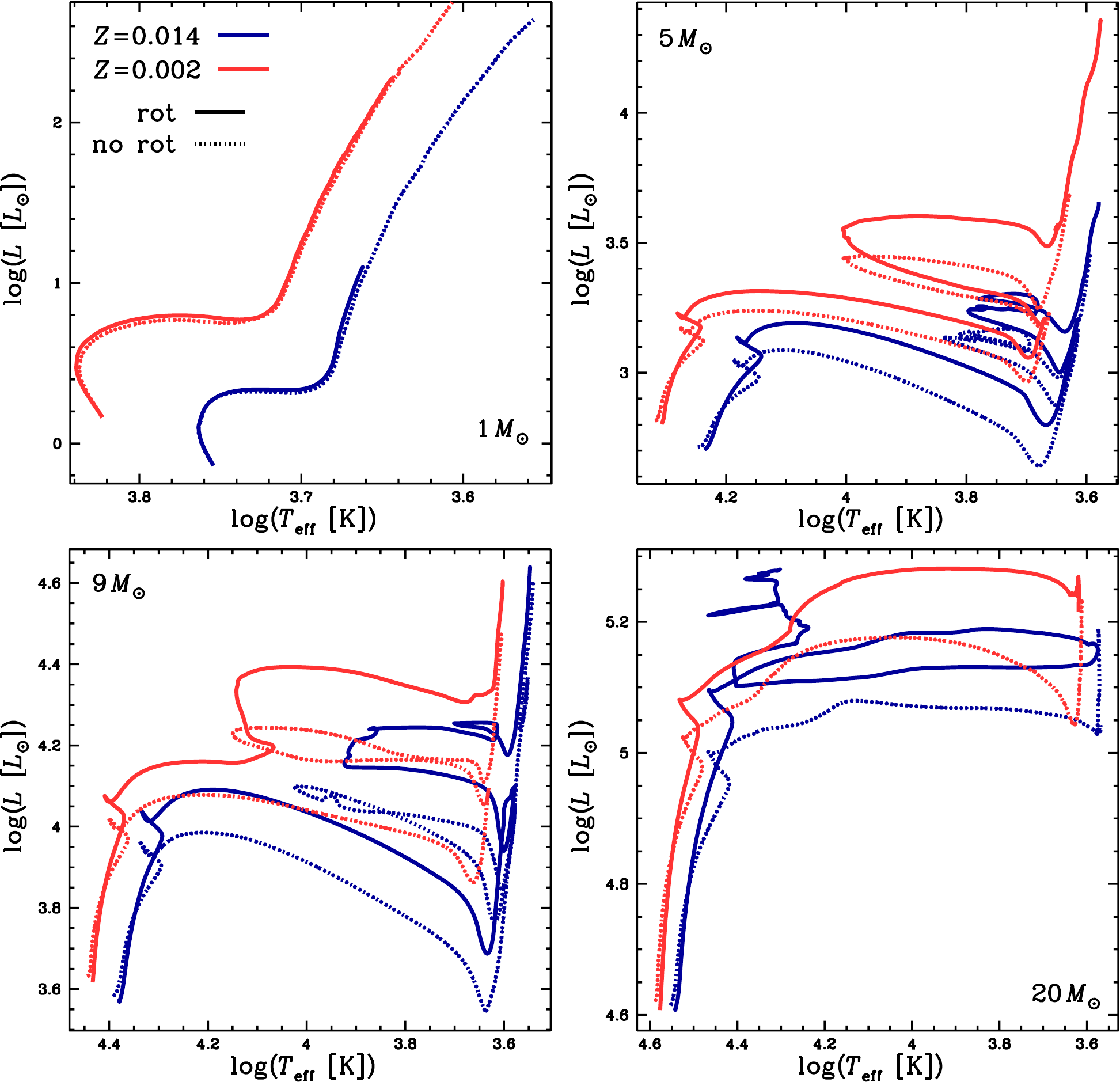}
\caption{Comparisons of evolutionary tracks with and without rotation for models with $M_\text{ini}=1$, $5$, $9$, and $20\,M_{\sun}$ at respectively $Z=0.002$ (red) and $0.014$ (blue). The solid lines are for rotating models, the dotted ones for non-rotating models.}
\label{Fig_compdhr}
\end{figure*}

Figure~\ref{Fig_compdhr} compares the tracks at $Z=0.002$ and $0.014$ of a few selected models ($M_\text{ini}=1$, $5$, $9$, and $20\,M_{\sun}$, dotted lines for the non-rotating models). As it is well known, low-metallicity tracks are shifted to the blue and to a higher luminosity with respect to tracks at higher $Z$. This behaviour is the result of the combination of two effects: first the opacity effect, and second the CNO effect. The opacity effect is the decrease of the opacity in the outer layers when the metallicity decreases. This favours a higher luminosity. At low $Z$, the lower initial CNO abundances force the star to reach a higher central temperature and a higher central density than at high $Z$ to get a given energy generation rate through CNO burning. This makes the star more compact.

The opacity effect is the main driver for the shift in effective temperature and luminosity of the $1\,M_{\sun}$. Indeed, for that mass, the abundances of the CNO elements do not have a significant effect on the energy generation since the energy is mainly produced through the pp-chains. In contrast, in the $20\,M_{\sun}$ model the opacity is dominated by the free electron scattering process, with only a slight dependence in the metallicity, while the CNO abundances play the most influential role.

For the $1\,M_{\sun}$ model, the change in metallicity does not produce a significant change in the radius. At $Z=0.014$, the radius on the ZAMS is $0.894\,R_{\sun}$, while it is slightly larger, $0.899\,R_{\sun}$ at $Z=0.002$. For the more massive models shown in Fig.~\ref{Fig_compdhr}, the stars on the ZAMS are more compact at $Z=0.002$ than at $0.014$. Typically, the radius of the $5\,M_{\sun}$ stellar model at that stage is equal to $2.42\,R_{\sun}$ at $Z=0.014$ and to $1.91\,R_{\sun}$ at $Z=0.002$, \textit{i.e.} $21\%$ smaller. For the $20\,M_{\sun}$ model, the corresponding radii are $5.30\,R_{\sun}$ at $Z=0.014$ and $4.56\,R_{\sun}$ at $Z=0.002$. The decrease amounts to $14\%$.

The size of the convective core depends also on the metallicity, and is regulated by two mechanisms. On the one hand, lower opacities allow the radiation to escape more easily, reducing the size of the convective core. On the other hand, a higher central temperature induces that the energy production is more concentrated near the core, favouring convection. On the ZAMS, all our models between $0.8$ and $20\,M_{\sun}$ at $Z=0.002$ have a larger core than at solar metallicity. Above $25\,M_{\sun}$, the initial size of the convective core is the same between both metallicities. During the subsequent evolution, the mass of the convective core is always larger at $Z=0.002$ than at $Z=0.014$ (for the high mass range, this is due to the fact that mass loss is much stronger at solar metallicity, tending to reduce the size of the core).

For the post-MS evolution, three features are worth noting: 
\begin{enumerate}
\item the blue loops appearing along the tracks for the $5$ and $9\,M_{\sun}$ are more extended at $Z=0.002$ than at $Z=0.014$, 
\item for a given initial mass, the loop occurs at a higher luminosity, 
\item the positions of the red giant branch and of the RSG are shifted towards the blue. All these features have already been obtained and discussed
in previous works \citep[][]{Schaller1992a,Stothers1973a,Girardi2000a}.
\end{enumerate}

\section{Properties of the rotating models \label{SecResultsROT}}

\subsection{Hertzsprung-Russell diagram and lifetimes\label{SubSecHRDROT}}

\begin{figure*}
\centering
\includegraphics[width=.9\textwidth]{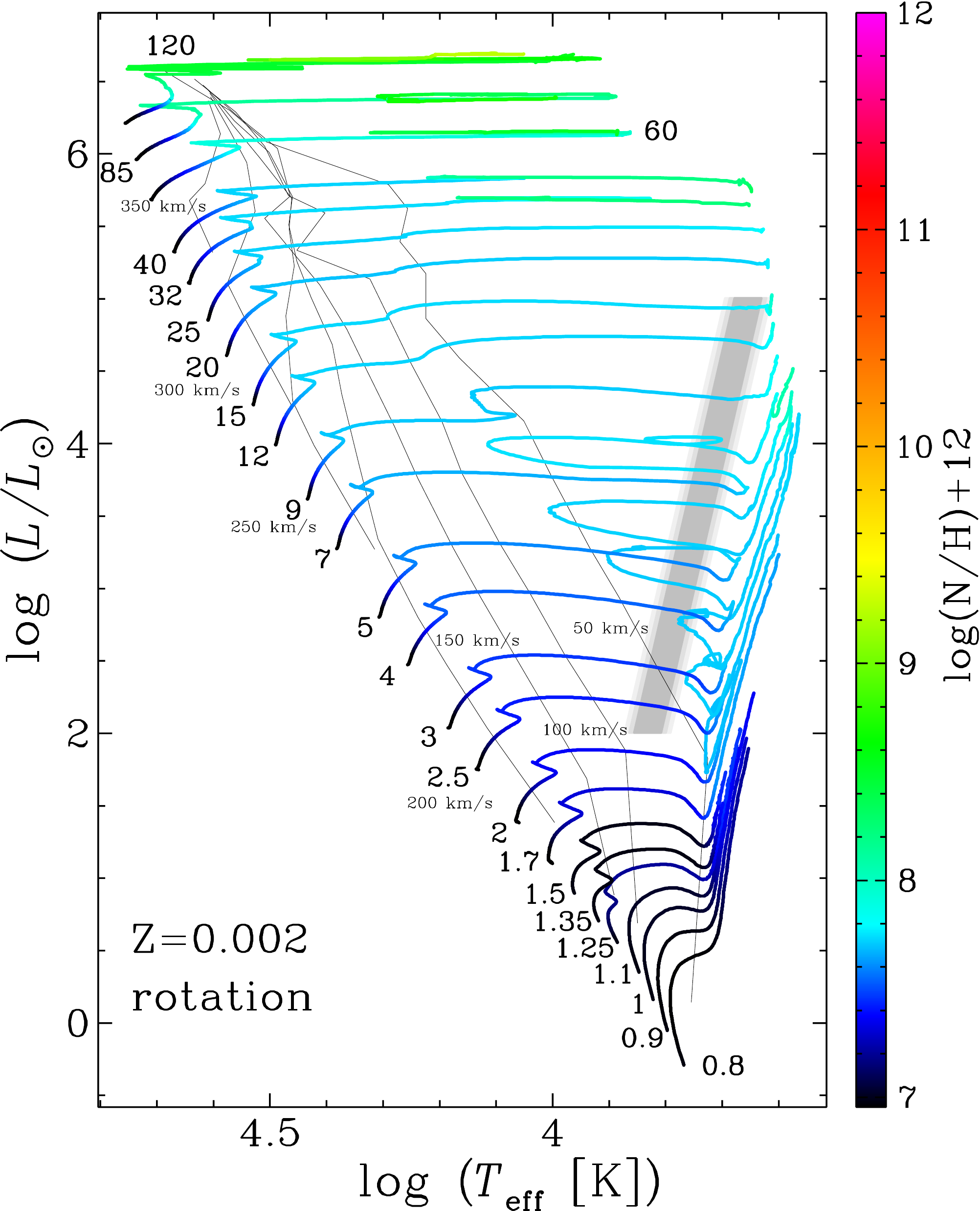}
\caption{Same as Fig.~\ref{Fig_HRDnorot} with the rotating models. Lines of iso-velocities are drawn through the diagram.}
\label{Fig_HRDrot}
\end{figure*}

\begin{figure}
\centering
\includegraphics[width=.48\textwidth]{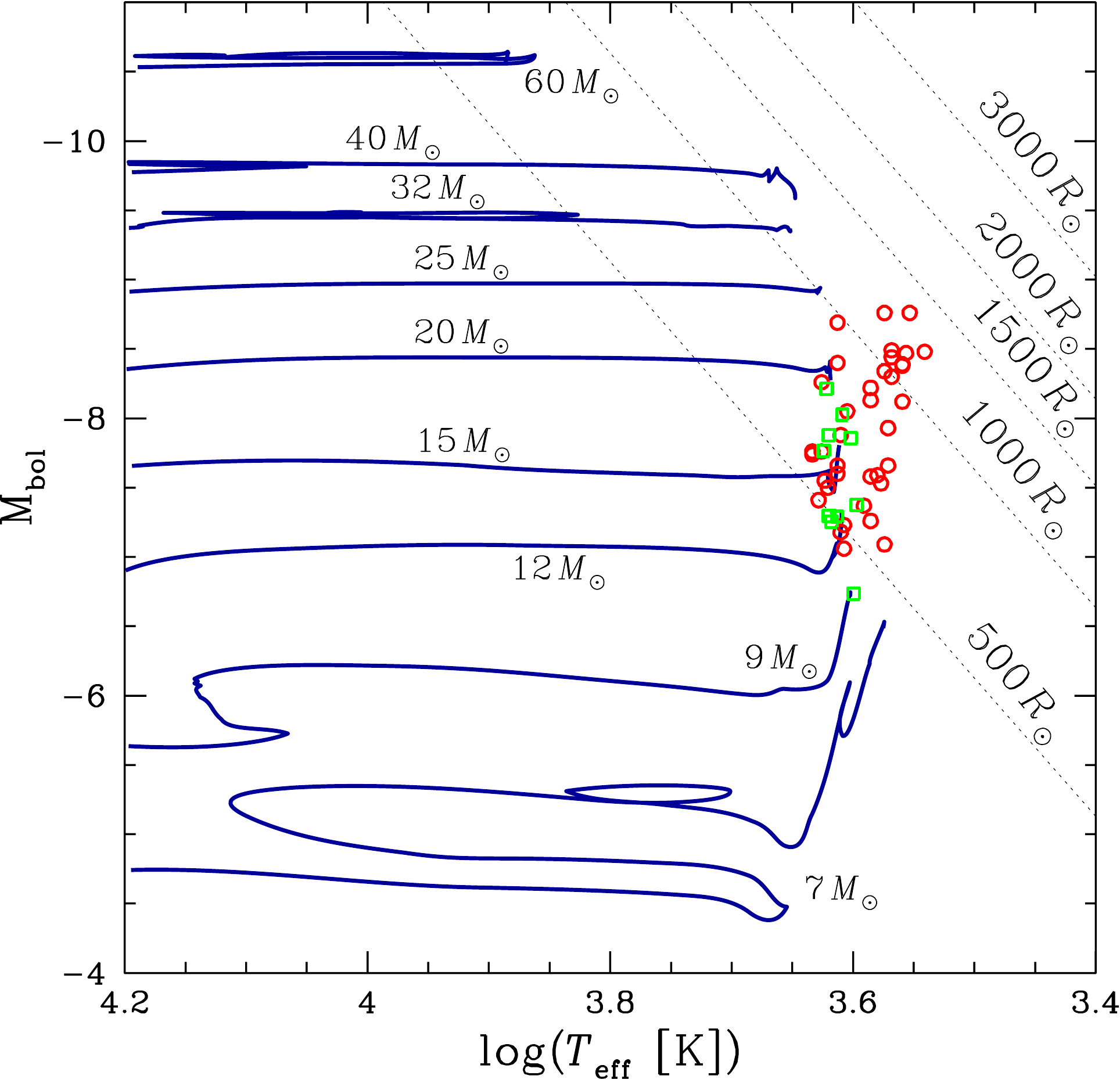}
\caption{Location of the RSGs  in the SMC from \citet{Levesque2006a} (red circles) and from \citet{Davies2013a} (green squares). The evolutionary tracks are the rotating models from the present work. The dotted lines shows some isoradius locations.}
\label{Fig_RSG}
\end{figure}

The tracks in the HRD for all the rotating models are presented in Fig.~\ref{Fig_HRDrot}, with lines of iso-surface velocities drawn across the diagram. Compared to Fig.~\ref{Fig_HRDnorot}, we note the following differences with respect to the non-rotating models:
\begin{itemize}
\item There is no longer any widening of the MS band occurring above $60\,M_{\sun}$ as in the non-rotating case. This is a consequence of the rotational mixing that prevents any significant redwards evolution during the MS phase by reducing the opacity in the outer layers. \textit{This shows that in this upper mass range ($M > 60\,M_{\sun}$), the MS width is very sensitive to rotational mixing, a sensitivity which was also found at solar metallicity}.
\item All models with $M_\text{ini}\leq40\,M_{\sun}$ evolve to the red part of the HR diagram (typically below $\log\left(T_\text{eff}\right) \sim3.7$). However the duration of the RSG phase for the $40$ and $32\,M_{\sun}$ is quite short, $2500$ and $1760\,\text{yr}$, respectively. The RSG lifetime for the $25\,M_{\sun}$ is $12600\,\text{yr}$, and $162500\,\text{yr}$ for the $20\,M_{\sun}$. Thus, the maximum luminosity of rotating RSGs is in the range $\log(L/L_{\sun}) =  5.3 - 5.5$, similar to non-rotating models. This value closely agrees with the upper limit given by \citet{Levesque2006a} as can be seen in Fig.~\ref{Fig_RSG}. The position of the RSG branch also well reproduces the observations from \citet{Davies2013a}.
\item Rotating models with $M_\text{ini}\gtrsim32\,M_{\sun}$ evolve back to the blue after a RSG phase or an LBV phase. The corresponding mass limit for non-rotating models is $40\,M_{\sun}$ so slightly above the rotating ones.
\item Extended blue loops, crossing the Cepheid instability strip occur for masses between $3$ and $7\,M_{\sun}$ (corresponding to a luminosity range between $2.75\lesssim\log(L/L_{\sun})\lesssim4$). The bluewards extension is similar or slightly wider than for the non-rotating models. The $9\,M_{\sun}$ rotating model seems to jump directly on the tip of a blue loop starting from a blue position in the HRD.
\item The tip of the red giant branch does not appear to be very different in the rotating models. We must stress here that not all the models have been computed strictly up to the He-flash, preventing us from reaching more firm conclusions. The above assessment relies on the case of the $1.7\,M_{\sun}$ model that was computed up to that stage.
\item Changes in the surface abundances of rotating models occur at a much earlier evolutionary stage than in non-rotating ones of similar initial mass. This point is discussed in more detail in Sect.~\ref{SubSecSurfAbund}. 
\end{itemize}

Figure \ref{Fig_compdhr} (already commented in Sect.~\ref{SecResultsNOROT}) compares also the rotating tracks (solid lines). We have to keep in mind that for the $1\,M_{\sun}$, the two models for the two metallicities start their evolution on the ZAMS with the same equatorial velocity ($50\,\text{km}\cdot\text{s}^{-1}$). For the more massive models shown in Fig.~\ref{Fig_compdhr}, the models start on the ZAMS with the same ratio $V/V_{\rm crit}=0.40$. At $Z=0.014$, this value implies an initial equatorial velocity equal to $219$, $248$ and $274\,\text{km}\cdot\text{s}^{-1}$ for the $5$, $9$, and $20\,M_{\sun}$ models, respectively. At $Z=0.002$, a slightly higher velocity is obtained: $228$ ($+4\%$) , $255$ ($3\%$) and $305$ ($11\%$) $\text{km}\cdot\text{s}^{-1}$ for the $5$, $9$, and $20\,M_{\sun}$, respectively.

For the initial velocity considered here, the track of the $1\,M_{\sun}$ is nearly unchanged by rotation at $Z=0.014$ and at $Z=0.002$. For the higher masses, we find that at the end of the MS phase, the rotating tracks reach a higher luminosity. This comes from the more massive helium core and the more He-rich envelope resulting of rotational mixing \citep[][and \citetalias{Ekstrom2012a}]{Meynet2000a,Brott2011a}. However, this result depends strongly on the prescription used for the treatment of rotation \citep[see][]{Chieffi2013a}. Figure~\ref{Fig_compconv} shows the evolution of the mass of the convective core as a function of time inside the $5$ and $9\,M_{\sun}$ models. Rotation increases the mass of the H-burning convective core and thus the MS lifetime. On the other hand, the mass of the He-burning core is less affected by rotation. The core He-burning lifetime is also barely changed by rotation. The comparison of the size of the convective core between the rotating models at $Z=0.002$ and $Z_{\sun}$ leads to the same conclusion as in the non-rotating case.

For the $5$ and $9\,M_{\sun}$ models at both metallicities, rotation does not change the extension in effective temperature of the blue loop, but shifts it to a higher luminosity. The rotating $20\,M_{\sun}$ model at $Z_{\sun}$ evolves back to the blue after the RSG stage, while  the similar model at $Z=0.002$ ends its evolution in the red (like the non-rotating one) because of the lower mass loss.

\begin{figure}
\centering
\includegraphics[width=.48\textwidth]{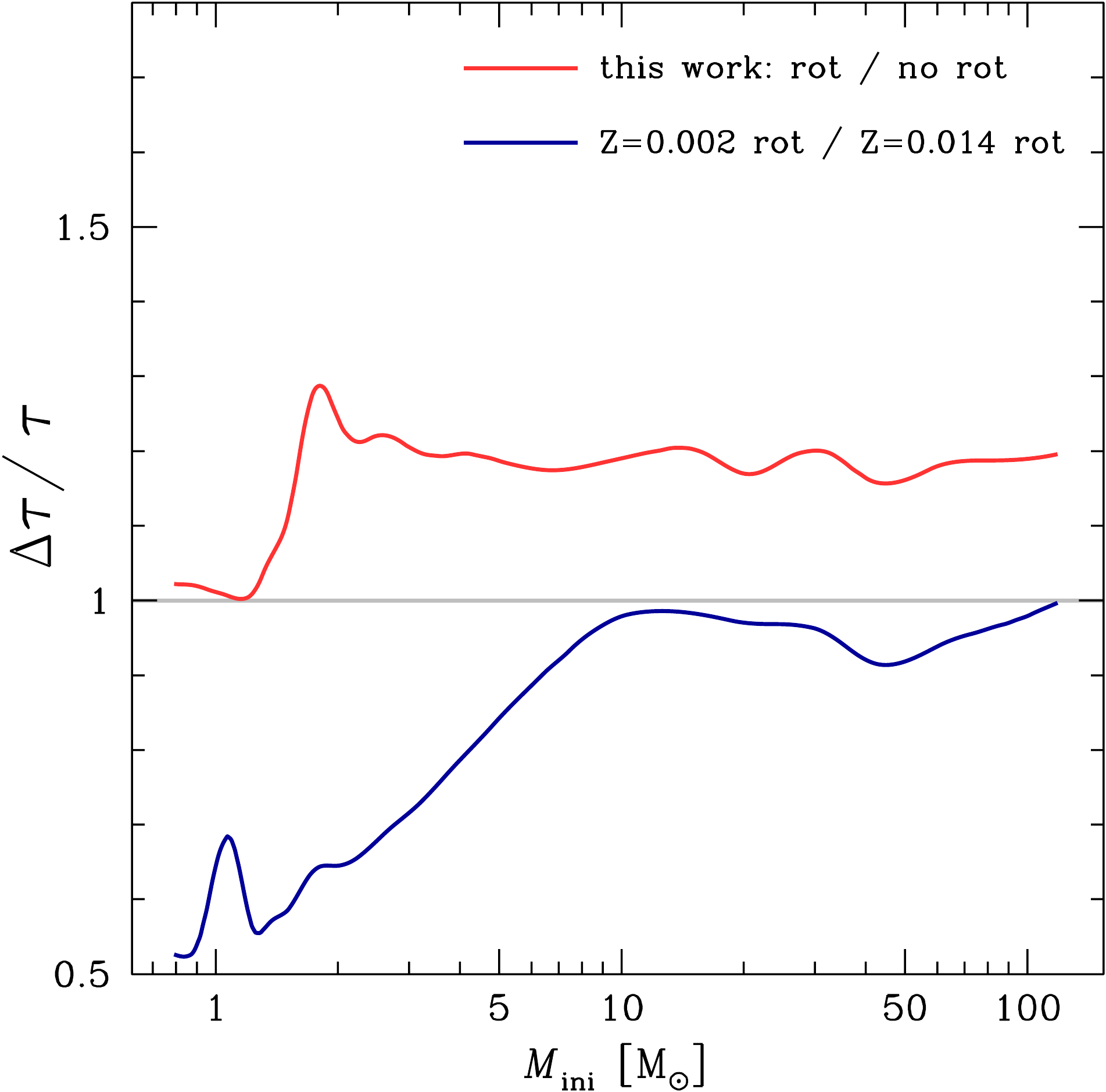}
\caption{Ratio of the MS lifetime durations. Comparison between the present work rotating and non-rotating models (red curve), and comparison between the present work's rotating and the \citetalias{Ekstrom2012a} models (blue curve).}
\label{Fig_tau}
\end{figure}

\begin{figure*}
\centering
\includegraphics[width=.98\textwidth]{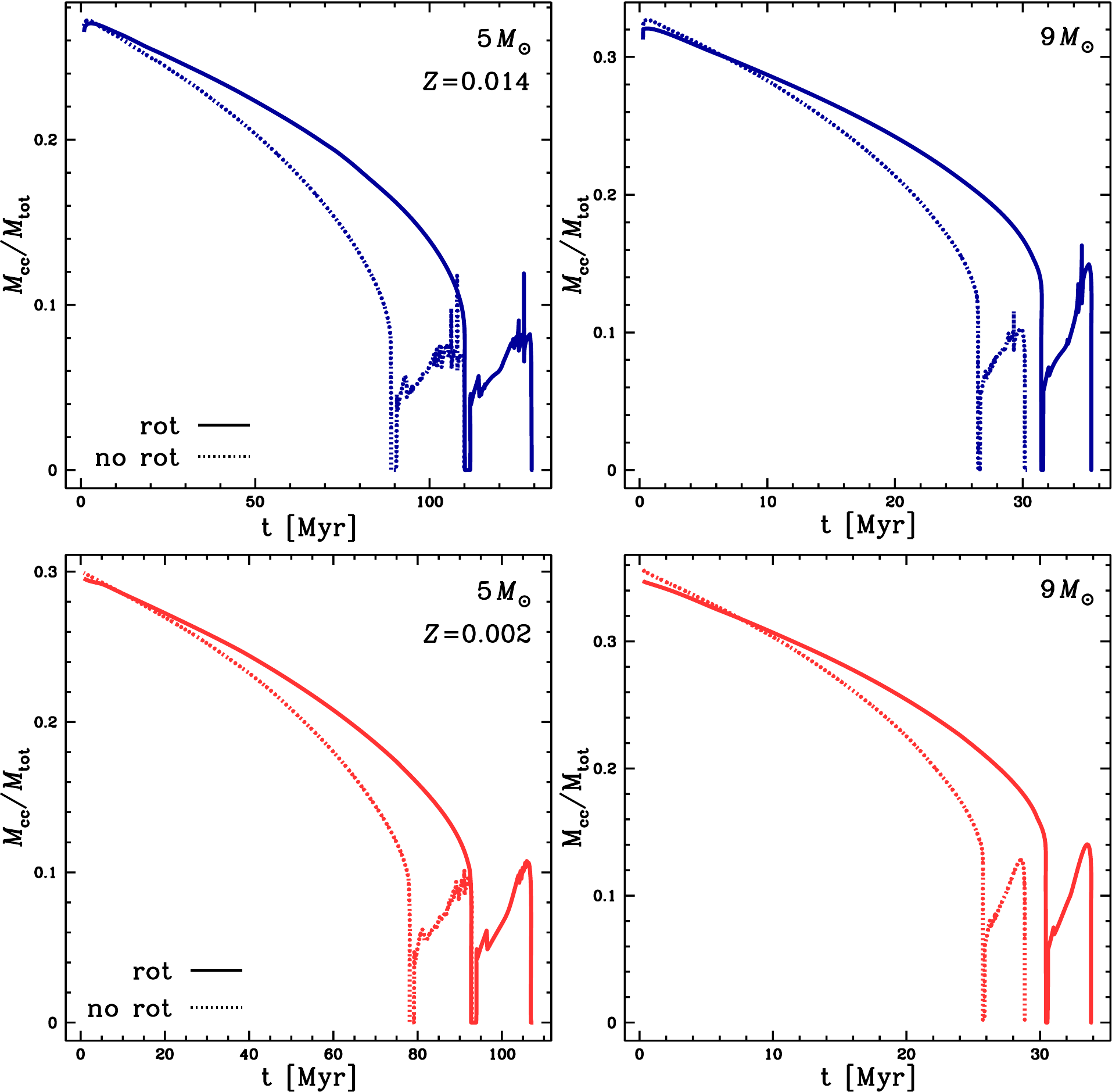}
\caption{Comparisons of the evolution of the mass of the convective core during the H- and He-burning phases inside the $5\,M_{\sun}$ (left panels) and $9\,M_{\sun}$ models (right panels). The solid (dotted) lines are for rotating (non-rotating) models, respectively. The upper panels show the $Z=0.014$ models, the lower ones the $Z=0.002$ models.}
\label{Fig_compconv}
\end{figure*}

As shown in Fig.~\ref{Fig_tau} (red solid line), an initial rotation of $V_\text{ini}/V_\text{crit} = 0.40$ enhances the time spent on the MS by about $20\%$. The increase in the MS lifetime is caused by rotational mixing, which refuels the core in fresh hydrogen. It is striking to see how relatively constant the increase in the MS lifetime appears to be over the whole mass range. This illustrates the fact that choosing the same value of $V_\text{ini}/V_\text{crit}$ for all the masses is equivalent to consider more or less similar strengths for the effects of rotation on all the initial mass models. This was already noted on the basis of the $Z=0.014$ models and is observed to be true at $Z=0.002$ too.

The MS lifetime for $M_\text{ini}\geq10\,M_{\sun}$ are nearly the same for both metallicities (see the blue line in Fig.~\ref{Fig_tau}). The models in this mass range have a slightly shorter MS lifetime at $Z=0.002$ than at $Z=0.014$ but the difference is in general less than $10\%$. The lifetime of lower initial mass models are in contrast significantly shorter at $Z=0.002$ than at $Z=0.014$. For instance, the MS lifetime of the $2\,M_{\sun}$ at $Z=0.002$ is decreased by about $35\%$ with respect to the corresponding model at $Z_{\sun}$. This results from the different  effects in various initial masses of a change of Z on the structure and on rotational mixing.

\begin{figure}
\centering
\includegraphics[width=.48\textwidth]{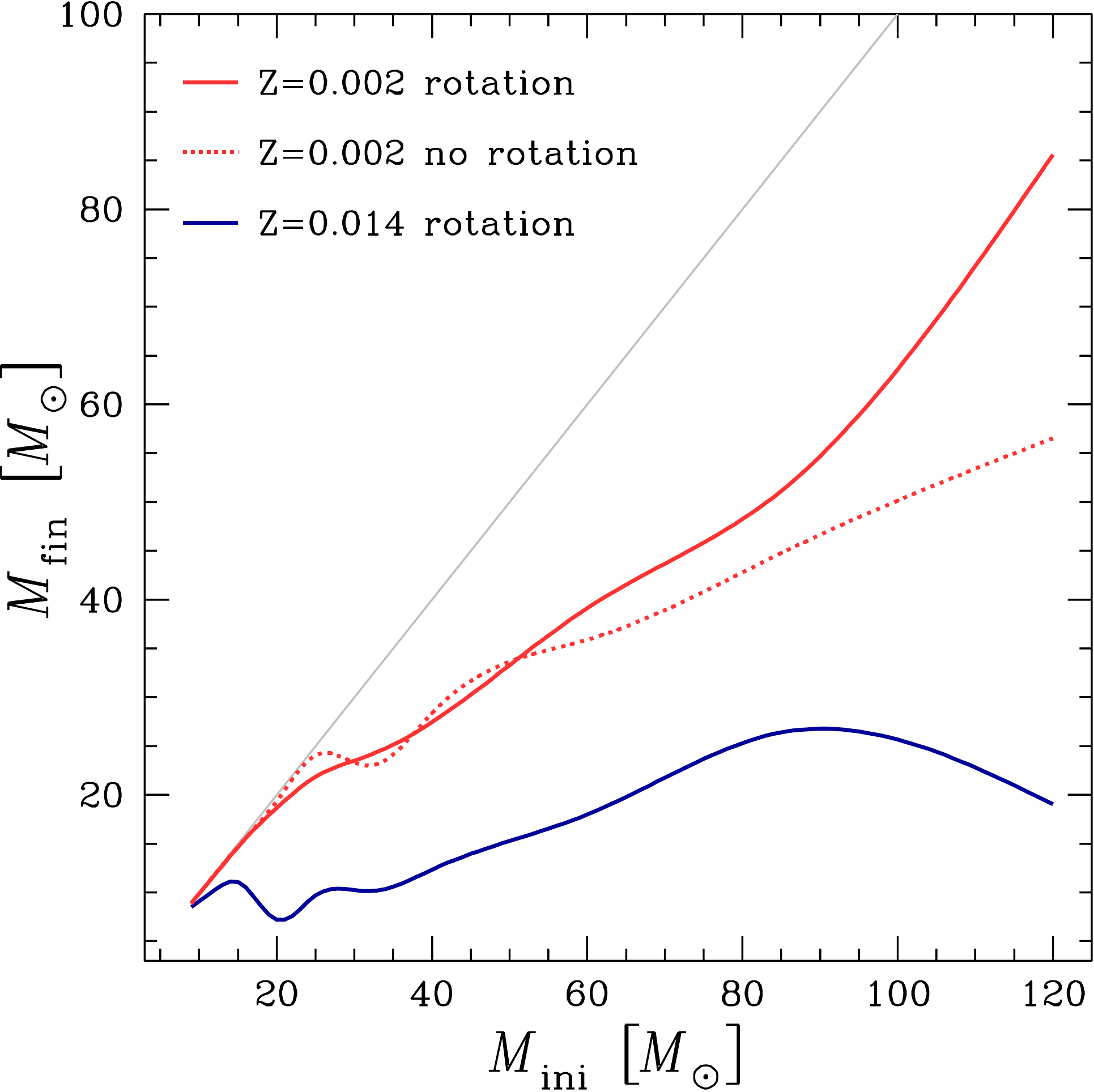}
\caption{Final mass \textit{vs} initial mass for models from $9$ to $120\,M_{\sun}$, present work's non-rotating models (red dotted line), rotating models (red solid line), and the rotating models from \citetalias{Ekstrom2012a} (blue line). The grey line corresponds to the hypothetical case without mass loss ($M_\text{fin} = M_\text{ini}$).}
\label{Fig_Mfin}
\end{figure}

Figure~\ref{Fig_Mfin} compares the final- to initial-mass relation for different sets of models. For $M_\text{ini}<50\,M_{\sun}$ at $Z=0.002$, the differences obtained between the rotating and non-rotating models are small. Above that mass, the rotating models up to $120\,M_{\sun}$ lose less mass than the non-rotating one. This might be surprising at first sight, since rotation is known to increase the mass-loss rates \citep{Pauldrach1986a,Friend1986a,Maeder2000a}. However this is correct when comparisons are made for identical positions in the HRD. When rotation is accounted for, the models remain for longer period in the blue part of the HRD, where the mass-loss rates are weaker. Indeed, by remaining in the blue part of the HRD, the rotating models pass from an O-type mass-loss rates directly to a WR mass-loss rates, without experiencing the bistability jump, the RSG, or the supra-Eddington mass-loss episodes. This was already the case at $Z=0.014$. At the lower metallicity studied here, the mass loss is negligible up to $\sim 20-25\,M_{\sun}$, due to the weaker stellar winds on the one hand, and to the longer presence in the blue part of the HRD compared to solar metallicity.

We see that the final mass at $Z=0.002$ for the rotating models lies between $40$ and $85\,M_{\sun}$ for an initial mass between $60$ and $120\,M_{\sun}$, respectively. The corresponding numbers at $Z=0.014$ are $18$ and $27\,M_{\sun}$. Therefore a decrease in $Z$ by a factor of $7$ produces an increase in the final mass by a factor of $2$ to $3$.

\subsection{Evolution of the surface velocities}\label{SubSecVelocities}

\begin{figure*}
\centering
\includegraphics[width=.45\textwidth]{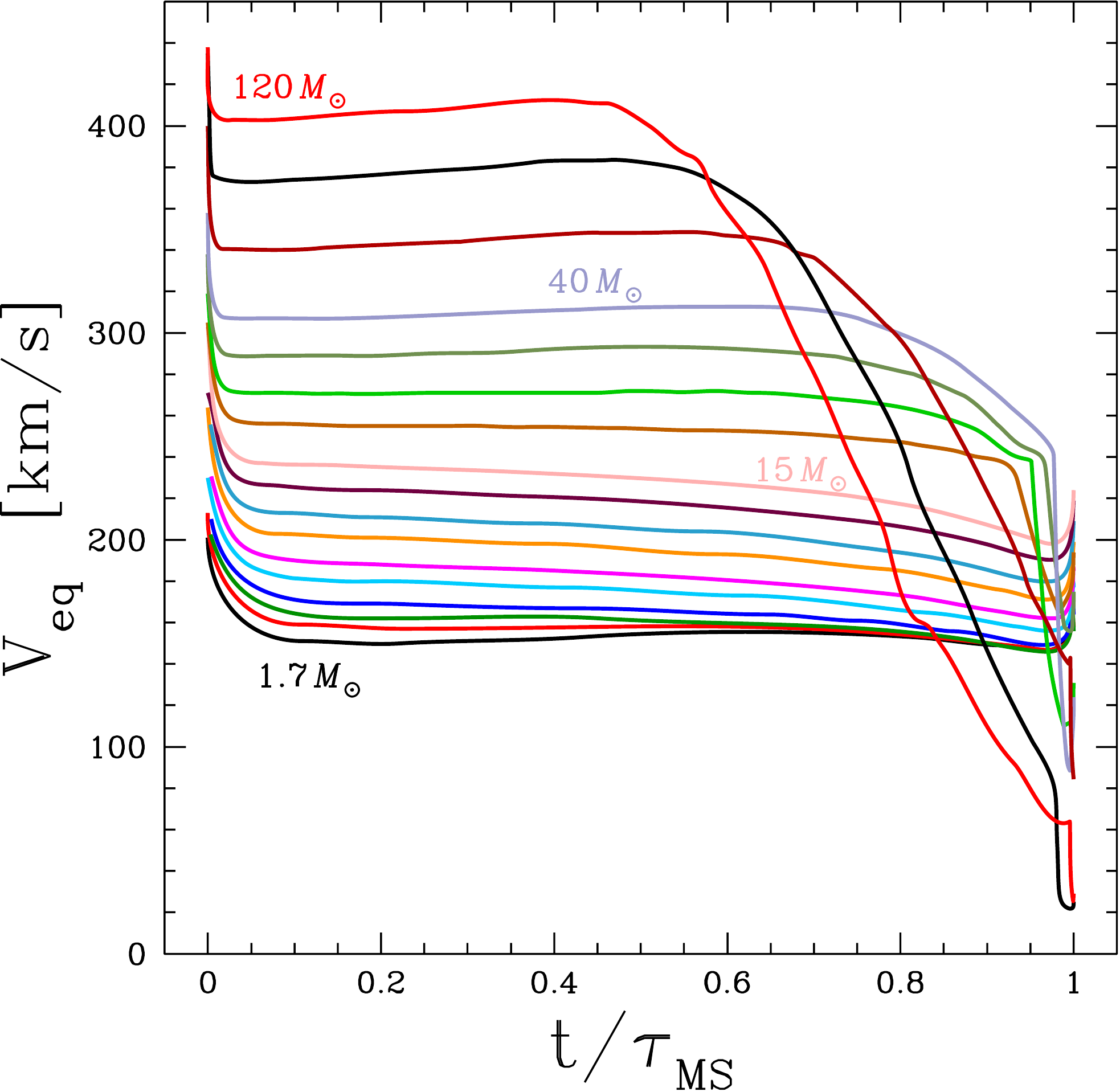}\hfill\includegraphics[width=.45\textwidth]{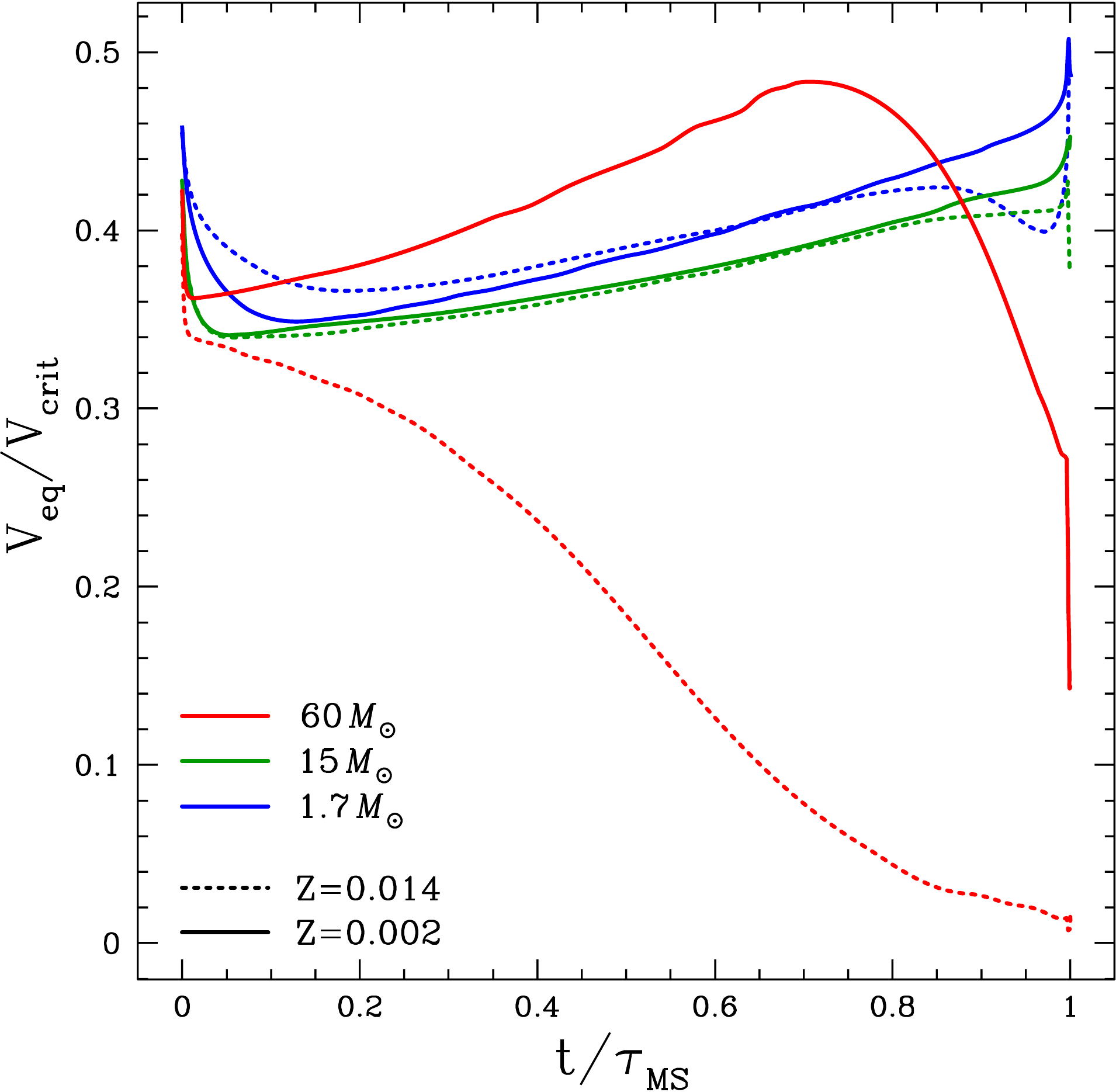}
\caption{Evolution of rotation on the MS as a function of the central mass fraction of hydrogen for all the stellar models from $1.7$ to $120\,M_{\sun}$. \textit{Left:} Evolution of the equatorial velocity. The curve for the $1.7\,M_{\sun}$ is the bottom black line at the beginning of the evolution, and that for the $120\,M_{\sun}$ the upper red line (increasing initial equatorial velocity for increasing mass). \textit{Right:} Evolution of the $V_\text{eq}/V_\text{crit}$ ratio for models with $M_\text{ini}=1.7$, $15$, and $60\,M_{\sun}$ at two different metallicities.}
\label{Fig_RotEvol}
\end{figure*}

The surface velocity and its evolution are depicted by the lines of iso-velocities drawn across the diagram in Fig.~\ref{Fig_HRDrot} and the data given in Table~\ref{TabListModels}. Figure~\ref{Fig_RotEvol} presents the evolution of the equatorial velocity (\textit{left}) during the MS for all models, and that of the $V_\text{eq}/V_\text{crit}$ ratio (\textit{right}) for a few selected models ($M_\text{ini}=1.7$, $15$, and $60\,M_{\sun}$) at $Z=0.002$ and $Z=0.014$. 

Since we computed models with identical initial $V_\text{eq}/V_\text{crit}$ ratio, the initial equatorial velocity varies with the mass considered: the higher the mass, the higher is the equatorial velocity needed to attain a given ratio. In Fig.~\ref{Fig_RotEvol} (\textit{left}), this is clearly visible: the $1.7\,M_{\sun}$ model draws the bottom black line, starting its MS evolution with $V_\text{eq}=201\,\text{km}\cdot\text{s}^{-1}$, while the most massive $120\,M_{\sun}$ model draws the upper red line on the ZAMS, starting its MS evolution with $V_\text{eq}=438\,\text{km}\cdot\text{s}^{-1}$. 

We bring our models onto the ZAMS assuming solid-body rotation, an assumption that is no longer made after the ZAMS. As soon as we allow for differential rotation, a quick readjustment of the $\Omega$-profile occurs inside the model, which explains the rapid drop in equatorial velocity at the very beginning of the evolution \citep{Denissenkov1999a}. One may wonder whether the solid-body assumption is justified and if that drop is in any way physical. To test this hypothesis, we computed the PMS evolution of a $9\,M_{\sun}$ at $Z=0.002$ with accretion and rotation. For this model, we adopt the same mass-accretion law as the one adopted in \citet{Behrend2001a} and we assume that the angular momentum of the accreted mass is the same as the angular momentum at the surface of the accreting object. We start from an initial hydrostatic core of $0.7\,M_{\sun}$ on the Hayashi line. Since at that stage the star is completely convective, we assume that the object rotates as a solid body. When through mass accretion $9\,M_{\sun}$ have accumulated onto the star, accretion is stopped and the PMS object evolves at constant mass towards the ZAMS. In this model we do not consider any disk-locking effect during that contraction phase. We choose the initial rotation of the accreting core in such a way that on the ZAMS the total angular momentum of the two models, \textit{i.e.} the one used as starting condition for the present grid and the one resulting from the modelling of the PMS phase, are the same.

\begin{figure}
\includegraphics[width=0.5\textwidth]{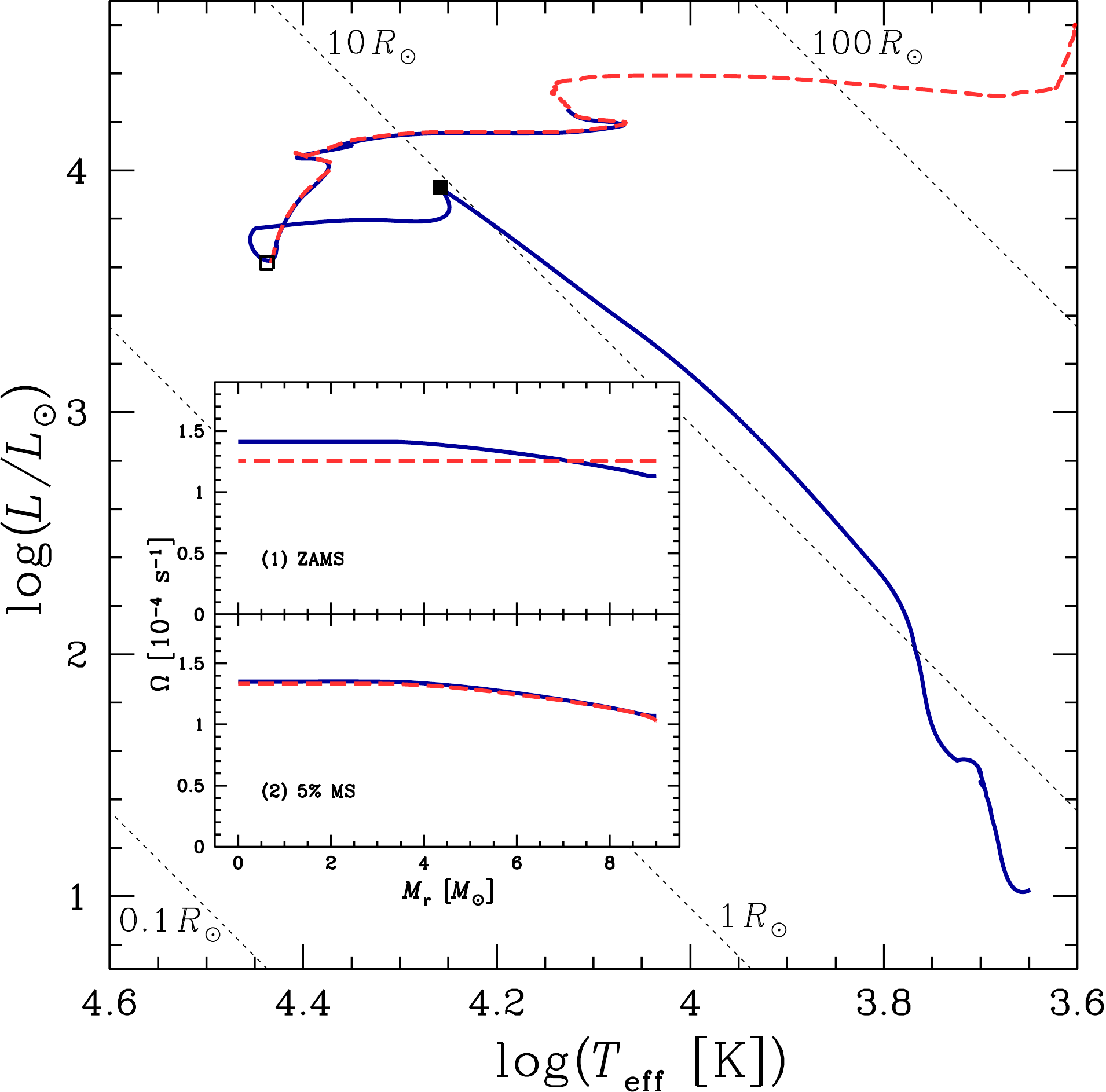}
\caption{Evolution of a rotating $9\,M_{\sun}$ model at $Z=0.002$ computed with a fully-consistent PMS evolution with accretion (blue solid line), compared to the track of the corresponding model of the present work (started on the ZAMS with a flat rotational profile, red dashed line). The black square shows the point where the accretion is stopped. The open square indicates the ZAMS. The dotted lines are isoradius lines. The inserted panel compares the $\Omega$-profile as a function of the Lagrangian mass inside the two models on the ZAMS (top) and after 5\% of the MS (bottom).}
\label{Fig_hrd_pms}
\end{figure}

The variation of $\Omega$ inside the two models is shown in the inset of Fig.~\ref{Fig_hrd_pms}. By construction the model used as a starting point for the present grid shows a flat rotational profile, while the one resulting from the modelling of the PMS phase shows a very shallow gradient, with a ratio of the central to surface angular velocity equal to $1.24$.

To ascertain however that the small differential rotation obtained in the model with the computed PMS does not affect the evolution during the MS phase, we pursued the evolution of this model during the MS phase. After a small duration corresponding to $5\%$ of the MS duration, we see that the rotation profile of the model started with a flat profile is already relaxed and does not differ significantly any more from the model computed with rotation during the whole PMS phase. The evolutionary tracks are very similar. The evolution of the surface abundances, the lifetimes, the evolution of the surface velocity and of the internal distribution of angular momentum are very similar in the two models. Concerning the drop in the surface velocity at the very beginning of the MS, the model computed with the fully-consistent PMS phase presents also a similar feature, showing that there is indeed a structural readjustment at the beginning of the MS. However, the drop is less deep in this model (9\%) than in the one brought on the ZAMS with solid-body rotation (20\%).

Thus this computation supports the view that starting with a solid body rotation on the ZAMS is well justified. It must however be emphasised that this question should be investigated for other initial masses, other values of the initial rotation, probably also for other accretion laws for the mass and the angular momentum \citep[][A\&A in press]{Haemmerle2013a}.

Once the equilibrium profile is reached, the equatorial velocity evolves under the action of the internal transport mechanisms (which tend to bring angular momentum from the contracting core towards the external layers, spinning up the stellar surface) and mass loss (which removes angular momentum from the surface, spinning down the stellar surface). 

The tracks can be classified in three families:
\begin{enumerate}
  \item the models with $M_\text{ini}$ between $20$ and $120\,M_{\sun}$ undergo a significant braking due to mass loss at the end of the MS phase. The higher the mass, the earlier the braking starts and the more efficient it is;
  \item the models with $M_\text{ini}$ between $1.7$ and $15\,M_{\sun}$ present no important change of their surface velocity once the first readjustment has operated. This illustrates, as already indicated above, that in the absence of strong stellar winds, the surface continuously receives angular momentum transported by the meridional currents, and to a smaller extent by shears. Were this not the case, the surface would slow down simply as a result of its inflation and the local conservation of the angular momentum;
  \item the models with $M_\text{ini}\leq1.5\,M_{\sun}$ have $\bar{V}_\text{MS}$ much lower than $V_\text{ini}$, but very similar to the velocity at the end of the MS phase. This is the result of the very rapid braking exerted by the magnetic fields at the beginning of the evolution.
\end{enumerate}
Note that none of the present models reach the critical velocity during their lifetime. At every moment the $V_\text{eq}/V_\text{crit}$ ratio remains below or around $0.5$ (see Fig.~\ref{Fig_RotEvol}, \textit{right panel}).

The comparison of the evolution of the $V_\text{eq}/V_\text{crit}$ ratio at different metallicities (see right panel of Fig.~\ref{Fig_RotEvol}) shows that for the $1.7$ and $15\,M_{\sun}$ models the change in metallicity does not bring any significant changes, while important differences, due to the effects of stronger mass-loss rates at $Z=0.014$ than at $Z=0.002$ are visible for the $60\,M_{\sun}$ models.

After the MS phase, the surface velocity rapidly decreases during the crossing of the Hertzsprung gap for all the models (see the lines of iso-velocities drawn across the diagram in Fig.~\ref{Fig_HRDrot} and the data given in Table~\ref{TabListModels}).

\subsection{Evolution of the surface abundances \label{SubSecSurfAbund}}

A comparison between Figs~\ref{Fig_HRDnorot} and \ref{Fig_HRDrot}, where the colour map indicates various levels of nitrogen surface enrichments, shows how rotation greatly extends the portions of the tracks where a nitrogen enhancement occurs. As already discussed in \citet{Maeder2009b}, the enrichment level depends both on the initial mass (higher at higher mass) and on the age (higher at more advanced stages). This trend is valid down to $1.7\,M_{\sun}$. Below this mass, magnetic braking, by enforcing a large gradient in the rotational velocity in the outer layers, boosts the mixing of the chemical species \citep[see also][]{Meynet2011a} and produces the peak seen in the low mass range in Fig.~\ref{Fig_NHtot}. As mentioned in \citetalias{Ekstrom2012a}, the magnetic braking might be overestimated in these low-mass models, since the resulting enrichment is higher than expected in this mass domain. A weaker magnetic braking would probably allow the models to closely fit both the surface velocity and the enrichment levels that are observed in stars of this mass range. The present models, however,  give the solution obtained when a solar magnetic braking law is applied to stars with $M_\text{ini}=1-1.5\,M_{\sun}$.

At the end of the MS phase, the $\left[\text{N}/\text{H}\right]$ are increased by more than a factor of $3$ with respect to the initial ratio at the surface of all rotating stars with $M_\text{ini} > 2.5\,M_{\sun}$, while in the corresponding non-rotating models, this ratio has kept the same value as on the ZAMS.

\begin{figure}
\centering
\includegraphics[width=.48\textwidth]{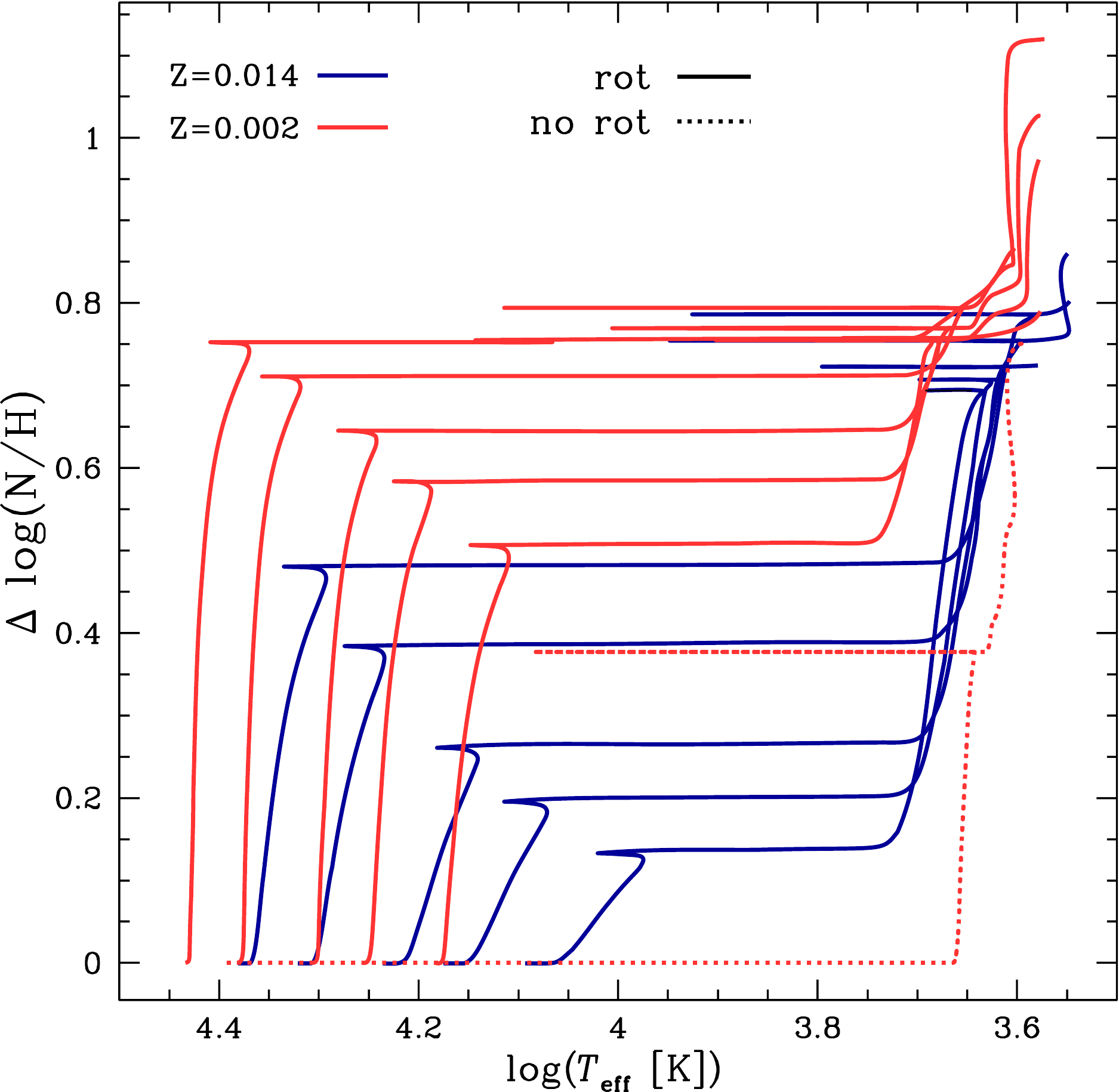}
\caption{Evolutionary tracks in the $\Delta\log(\text{N}/\text{H})$ {\it vs} $\log\left(T_\text{eff}\right)$ plane, with $\Delta (\text{N}/\text{H})=\log\left(\text{N}/\text{H}\right)_t-\log\left(\text{N}/\text{H}\right)_\text{ini}$, where $(\text{N}/\text{H})_t$ is the ratio of the nitrogen to hydrogen abundance (in number) at the surface of the star at the time $t$. The tracks are from models with $M_\text{ini}=3$, $4$, $5$, $7$, and $9\,M_{\sun}$ (from right to left). The solid red lines are for the models at $Z=0.002$ and the solid blue lines correspond to the models at $Z=0.014$ from \citetalias{Ekstrom2012a}. The dotted red line correspond to the non-rotating $7\,M_{\sun}$ model at $Z=0.002$.}
\label{Fig_NHteff}
\end{figure}

In Fig.~\ref{Fig_NHteff}, we compare the surface nitrogen enrichment in models with $M_\text{ini}=3-9\,M_{\sun}$ at $Z=0.002$ and $0.014$. The nitrogen enrichment is higher at $Z=0.002$ than at $Z=0.014$, a result that is in conformity with previous works \citep[][]{Meynet2000a,Brott2011a}. 

\begin{figure}
\centering
\includegraphics[width=.48\textwidth]{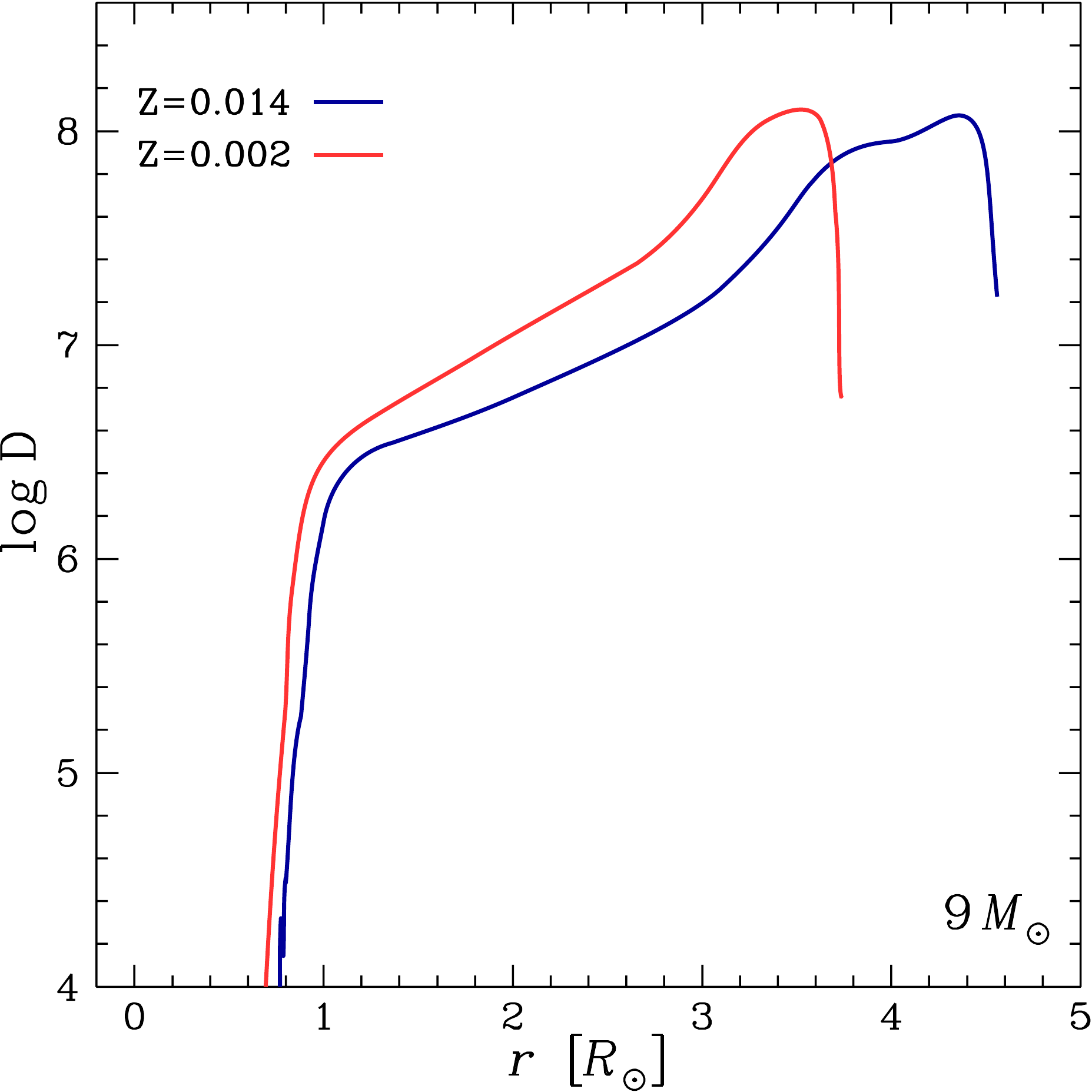}
\caption{Variation of the shear diffusion coefficient inside the $9\,M_{\sun}$ at the middle of the MS phase. The red curve is for the model at $Z=0.002$ and the blue one for the models at $Z=0.014$ \citepalias[from][]{Ekstrom2012a}.}
\label{Fig_compD}
\end{figure}

The models have the same value of $V_\text{ini}/V_\text{crit}=0.4$, so the higher enrichment comes from the mixing being more efficient of  at low metallicity. The causes are twofold:
\begin{itemize}
\item the stars are more compact at low $Z$, thus for similar values of the diffusion coefficients, the mixing timescale is shorter since it scales with the square of the radius;
\item the shear diffusion coefficient, which is the main responsible for the mixing of chemical species, is higher at a given radius in the models at $Z=0.002$ than at $Z=0.014$ (see Fig.~\ref{Fig_compD}).
\end{itemize}

\begin{figure}
\centering
\includegraphics[width=.48\textwidth]{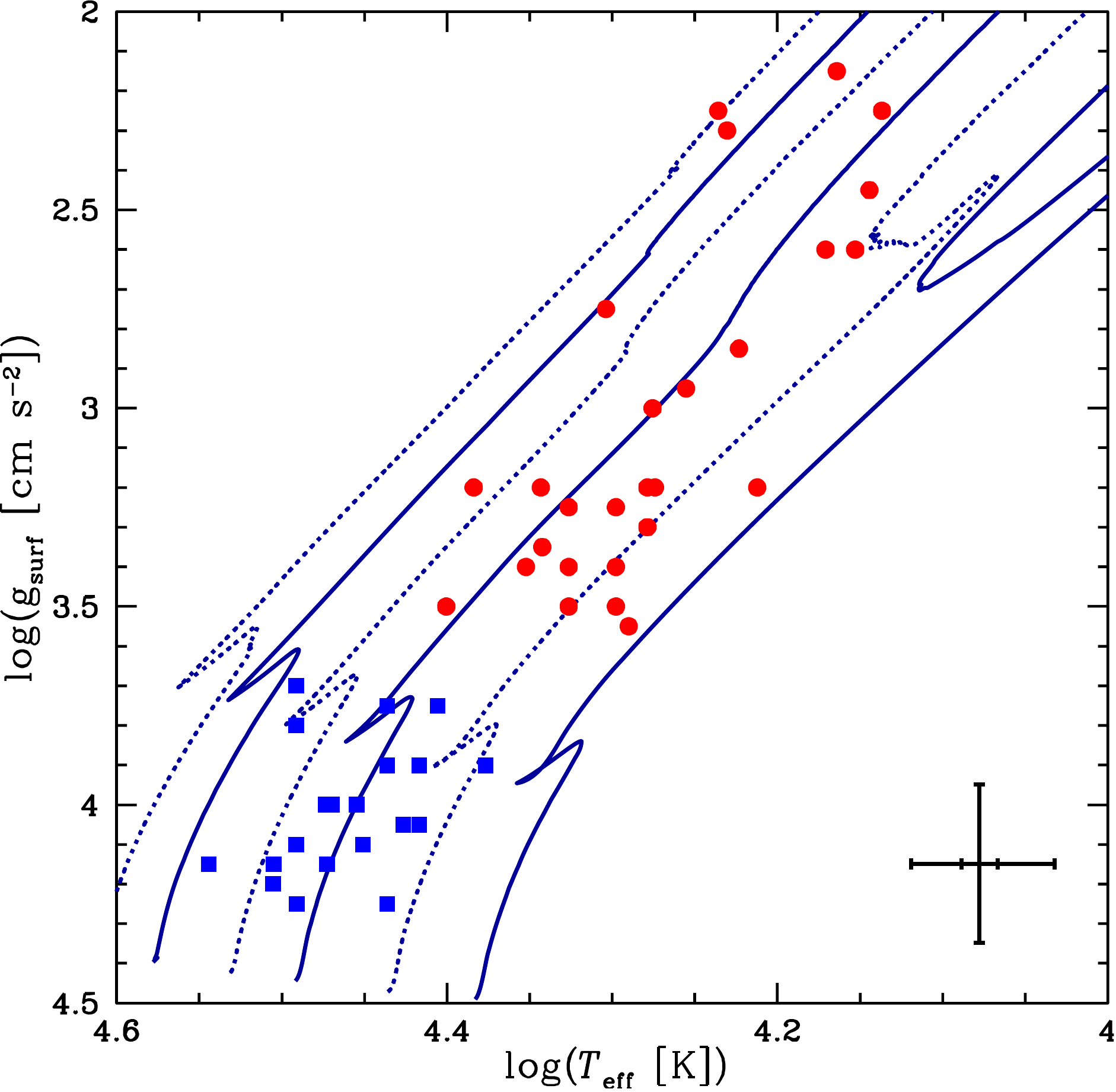}
\caption{Evolutionary tracks in the $\log(g)$ {\it vs} $\log\left(T_\text{eff}\right)$ plane for the $7$, $9$, $12$, $15$, $20$, and $25\,M_{\sun}$ rotating models (from right to left) at $Z=0.002$. The observational points are SMC stars observed by \citet{Hunter2007a} (blue squares: MS stars; red circles: post-MS stars). Typical errorbars are indicated on the bottom-right corner: $0.2\,\text{dex}$ in $g_\text{surf}$ and $1000\,\text{K}$ in $T_\text{eff}$ (corresponding to the smallest error shown at $\log(T_\text{eff}) = 4.6$ and to the biggest one at $\log(T_\text{eff}) = 4.0$.}
\label{Fig_gteff}
\end{figure}

\begin{figure}
\centering
\includegraphics[width=.48\textwidth]{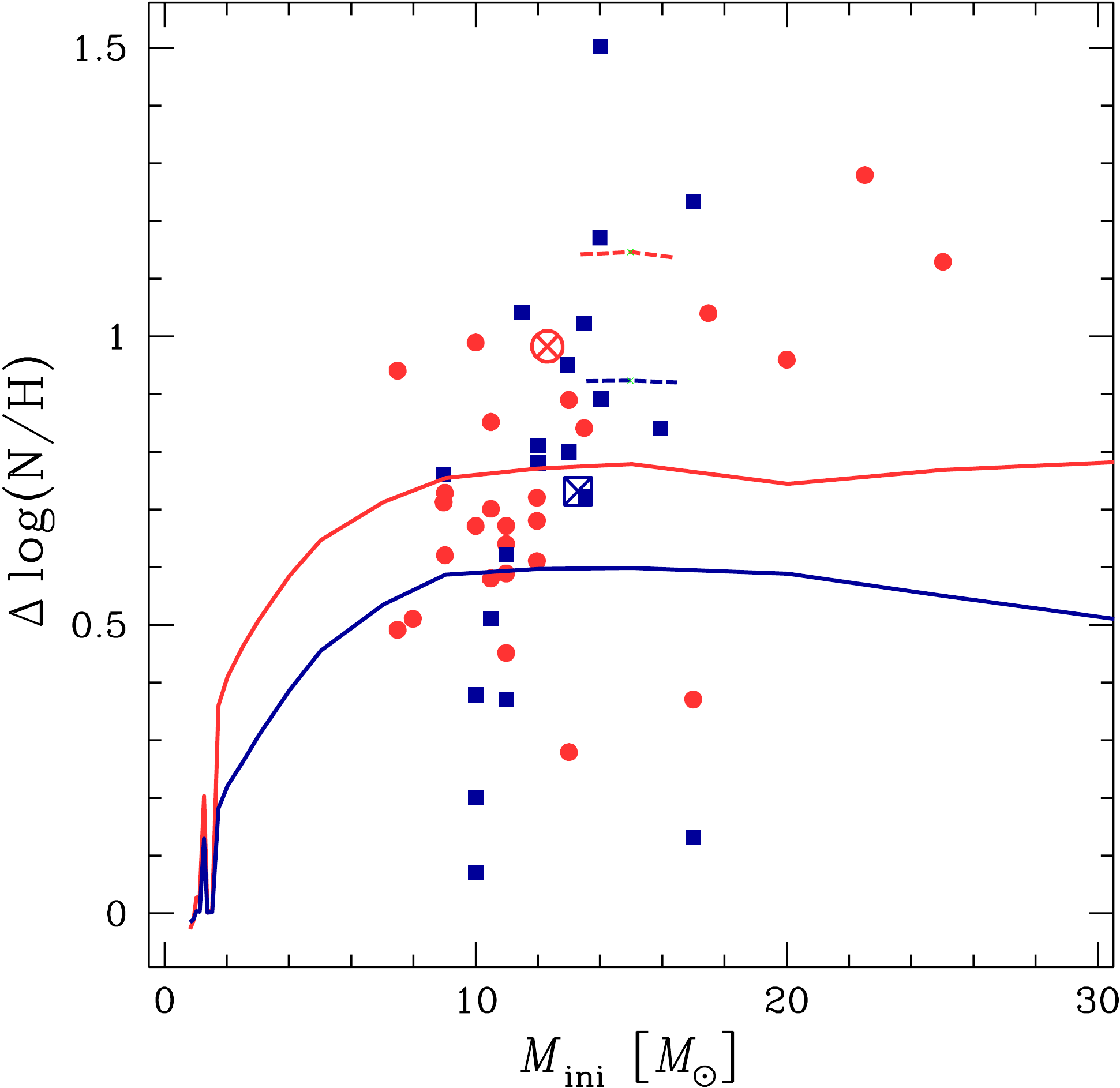}
\caption{Variation of the nitrogen enrichment as a function of the initial mass for the present rotating models (blue solid line: middle of the MS; red solid line: end of the MS). The short dashed lines show the value of $\Delta\log\left(\text{N}/\text{H}\right)$ for a $15\,M_{\sun}$ model when initial abundances for the SMC are used (see text). The observational points are from the same stars as in Fig.~\ref{Fig_gteff}. The big cross in square (circle) shows the mean observed values for the MS (post-MS) stars. The error bars on the values of $\Delta \log\left(\text{N}/\text{H}\right)$ are between $0.2$ and $0.3\,\text{dex}$ \citep[see Table 2 in][]{Hunter2009a}.}
\label{Fig_NHtot}
\end{figure}

The enrichment obtained in the present models can be compared with the one observed by \citet{Hunter2009a} at the surface of B-type stars in the SMC. Figure~\ref{Fig_gteff} present all the stars listed in their Table 2 that are observed in the directions of the two SMC clusters NGC 346 and NGC 330, and for which an estimate for the nitrogen to hydrogen ratio is given. We did not consider here the stars for which only an upper limit for this quantity is given, since they represent a poor constraint on the models. Using Fig.~\ref{Fig_gteff}, we assign to each observed star a value for its initial mass. Using that value, we plot in the plane $M_\text{ini}$ versus $\Delta\log\left( \text{N}/\text{H} \right)$\footnote{$\Delta\log\left( \text{N}/\text{H} \right)$ is the logarithm of the N/H ratio normalised to the initial one.} the position of the observed stars (Fig.~\ref{Fig_NHtot}) and compare them with the predictions of the present models. Note that the enrichment obtained at the surface of the $Z=0.002$ models do \textit{not} result from a calibration of the diffusion coefficients. The enrichment is the one predicted by applying the same physics as the one used for the $Z=0.014$ models (and calibrated\footnote{The $Z_{\sun}$ models were calibrated in the sense that they were computed with a shear diffusion coefficient set for reproducing the averaged observed chemical enrichments of B-type stars rotating with an average surface velocities.} for this metallicity only). 

When comparing the theoretical curves with the observation points, a preliminary remark must be made: our  models have been computed with scaled solar abundances for the heavy elements. This means that the initial values for the ratios $\text{C}/\text{N}$ and $\text{O}/\text{N}$ (in mass fractions) are $3.5$ and $8.7$, respectively. According to \citet{Hunter2009a}, the corresponding observed ratios of the HII regions in the SMC (expected to be equal to the composition of the young stars) is $3.7$ and $33.0$, respectively. In the case of a complete processing of carbon and oxygen into nitrogen, the maximum value of  $\Delta\log\left(\text{N}/\text{H}\right)$ in our models would be $1.12$, while it would be $1.6$ in the observed stars. Of course these are extreme values in the sense that to be obtained, the material should have undergone a complete CNO processing. The data points we compare our models to do not correspond to such an extreme case, since the material processed in the stellar interior is partially mixed with that of the envelope, whose composition has not been affected by the CNO cycle. But we expect that the enrichment we obtain in our models underestimate the one obtained in the SMC stars, because of this difference in the initial composition. 

We now focus on the subset of stars for which the position in the $\log(g)$ versus $\log\left(T_\text{eff}\right)$ plane identifies them as MS stars (the blue squares in Figs.~\ref{Fig_gteff} and \ref{Fig_NHtot}). We computed for the observed sample an average initial mass $\langle M_\text{ini}\rangle=13.3\,M_{\sun}$, an averaged $\langle\Delta\log\left(\text{N}/\text{H}\right)\rangle=0.73$, and an averaged $\langle V_\text{eq}\rangle=\langle V\sin(i)\rangle \cdot 4/\pi=100\,\text{km}\cdot\text{s}^{-1}$. For the same initial mass, our models obtain $\langle\Delta\log\left(\text{N}/\text{H}\right)\rangle=0.60$ for $\langle V_\text{eq}\rangle\simeq220\,\text{km}\cdot\text{s}^{-1}$. If we do the same exercise for stars beyond the MS phase, we obtain for the observations $\langle M_\text{ini}\rangle=12.3\,M_{\sun}$, $\langle\Delta\log\left(\text{N}/\text{H}\right)\rangle=0.98$, and $\langle V_\text{eq}\rangle=60\,\text{km}\cdot\text{s}^{-1}$. The models give for this initial mass $\langle\Delta\log\left(\text{N}/\text{H}\right)\rangle=0.77$, and $\langle V_\text{eq}\rangle$ between $50$ and $100\,\text{km}\cdot\text{s}^{-1}$ (see Fig.~\ref{Fig_HRDrot}). Thus the present models provide nitrogen enrichments which are slightly below the observed ones, but at such a level that the difference can be explained by the differences in the assumptions made for the initial compositions and/or velocity. To check this point, we computed a rotating $15\,M_{\sun}$ with $V_\text{ini}/V_\text{crit} = 0.4$ using for the initial composition the one measured for the SMC \citep[see for instance][]{Hunter2009a,Brott2011a,Brott2011b}. In that case the values of $\Delta\log\left(\text{N}/\text{H}\right)$ obtained at the middle and the end of the MS are $0.92$ and $1.15$, respectively, \textit{i.e.} they are shifted by slightly more than a factor of two with respect to the  enrichment obtained at the same stage when scaled solar initial abundances for the heavy elements are used. Using smaller initial rotations, more in line with the observed velocities in the sample of \citet{Hunter2009a} should provide a reasonable fit to the observations.

\subsection{Blue and red evolution \label{SubSecMassiveEvol}}

\begin{figure*}
\centering
\includegraphics[width=\textwidth]{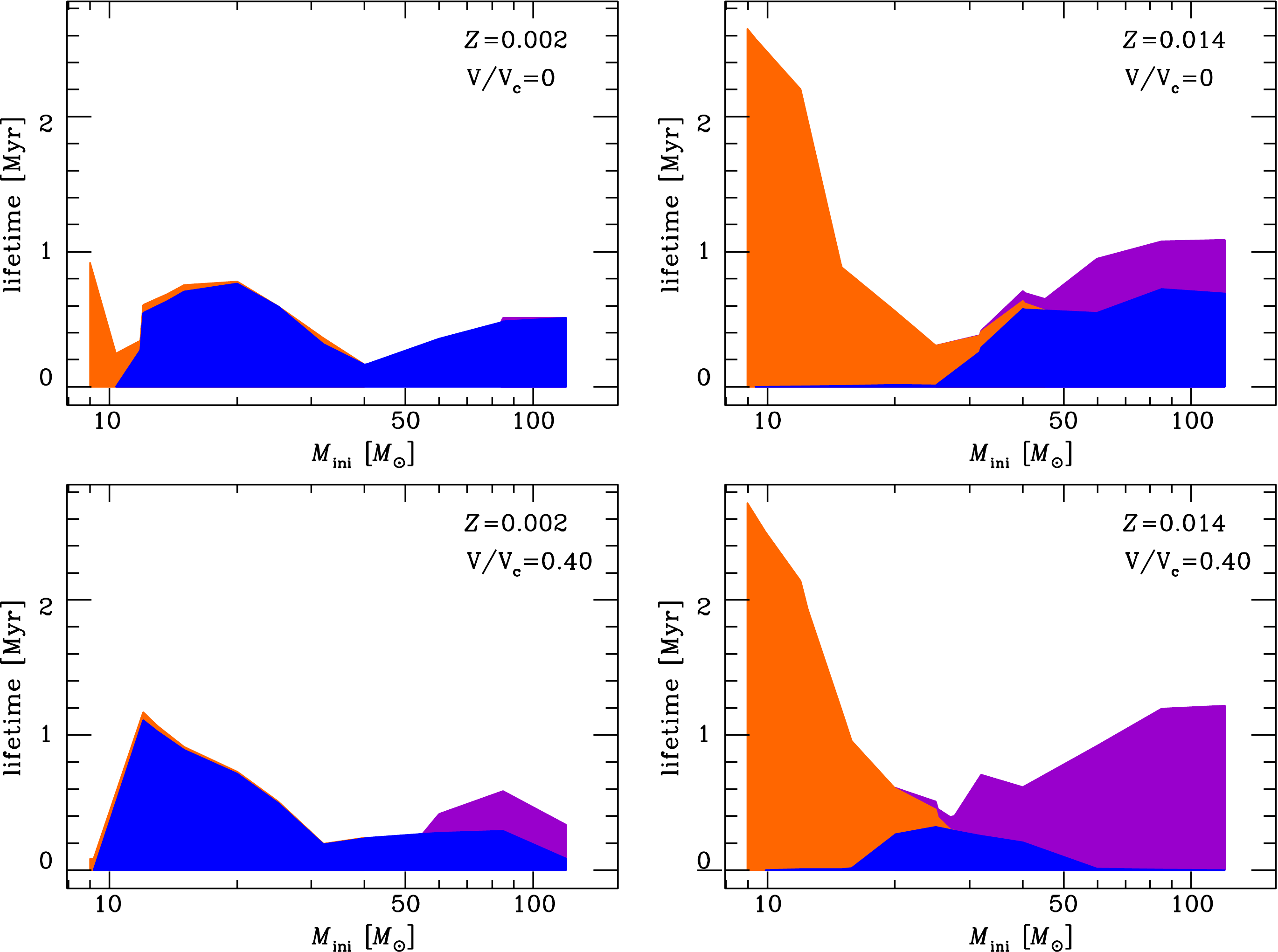}
\caption{Durations of the BSG (blue region) and RSG (red region) phases as well as of the WR phase (purple region) for stars as a function of the initial mass. Left panels: $Z=0.002$; right panels: $Z=0.014$. Top panels: non-rotating models; bottom panels: rotating models.}
\label{Fig_timeBR}
\end{figure*}

We defer a detailed discussion of massive star populations expected from our models to a forthcoming paper. Here we just want to emphasise some important differences between the models at $Z=0.002$ and $0.014$ in the context of the blue or red evolution. In Fig.~\ref{Fig_timeBR}, the durations of the red and blue supergiant (BSG) phases are shown for different stellar models. For the most massive stars, the durations of the Wolf-Rayet (WR) phase are also shown\footnote{Note that at $Z=0.002$, some evolved stars does not fulfil the criteria to be one of the plotted type of star, and are thus not represented in this plot. They can be yellow supergiants, or any other type of supergiant stars.}. To build this plot, we count as RSGs all the post-MS models with an effective temperature $\log\left(T_\text{eff}\right) \leq 3.66$ \citep[see \textit{e.g.}][]{Eldridge2008a} and an initial mass $\geq 9\,M_{\sun}$, as BSGs all the models with a $\log(T_\text{eff}) \geq 4.15$ and a surface gravity $\log(g) > 2.75$, and as WR all the models with a surface hydrogen mass fraction  inferior to $0.3$ and a $\log\left(T_\text{eff}\right) \geq 4.0$. Let us mention that these definitions may differ from those that we would obtained on the basis of spectral features. For instance, WR features may appear in the spectra of stars that have surface hydrogen abundance still higher than $0.3$. These remarks have to be kept in mind when comparisons are made with observations. Here however we only focus on theoretical aspects.

The main striking facts when looking at Fig.~\ref{Fig_timeBR} are the following:
\begin{enumerate}
\item At $Z=0.002$, our stellar models with $M_\text{ini}\lesssim20\,M_{\sun}$ spend most of their core He-burning lifetime as BSGs, while at $Z=0.014$, most of the core He-burning phase is spent in the RSG phase. This is true for rotating and non-rotating models. The question of the blue to red evolution is still a very debated point in the literature \citep[see \textit{e.g.}][]{Renzini1992a,Stancliffe2009a,Meynet2013a}. We revisit that question below studying the impact of various prescriptions for rotation.
\item Stars passing through a RSG stage before entering in the WR region only appear at $Z=0.014$ in a mass range between $32$ and $50\,M_{\sun}$ for non-rotating models, and between $20$ and $32\,M_{\sun}$ for models with $V_\text{ini}/V_\text{crit}=0.4$.
\item For both metallicities, rotation lowers the inferior mass limit for single stars to become WR stars and increases the WR lifetimes as was already discussed in previous works \citep[][]{Meynet2005a,Georgy2009a,Georgy2012b,Chieffi2013a}.
\end{enumerate}

The first point above is in sharp contrast with a result we obtained previously. In \citet[hereafter MM01]{Maeder2001a}, we computed models for $Z=0.004$, only for stars in a limited mass range between $9$ and $60\,M_{\sun}$. The non-rotating models are qualitatively similar to those presented here. In contrast, rotating models with initial masses between $9$ and $20\,M_{\sun}$ from \citetalias{Maeder2001a} show a nearly equal share between a blue and red position in the HRD. This is very different from the present results, and the reason for this comes from the differences in the choice we made for the shear diffusion coefficient. The shear diffusion coefficient used for computing the present models is higher in regions with mean molecular weight gradients than the one used to compute the \citetalias{Maeder2001a} models (both expressions give similar results in regions with no abundance gradient).

\begin{table}
\caption{Summary of the properties of the two models used to discuss the position of the star in the HRD during the central He-burning phase.}
\label{Table_HRD_HeBurning}
\begin{tabular}{c|ccc|c}
\rule[-2.5mm]{0mm}{4mm}Name & $M_\text{ini}\,[M_{\sun}]$ & $Z$ & $\frac{V_\text{ini}}{V_\text{crit}}$ & $D_\text{shear}$\\
\hline
\citetalias{Maeder1997a} & $15$ & $0.002$ & $0.5$ & \citet{Maeder1997a}\\
\citetalias{Talon1997a} & $15$ & $0.002$ & $0.5$ & \citet{Talon1997a}\\
\hline
\end{tabular}
\end{table}

Why would a stronger shear mixing in regions with an abundance gradient change so much the duration of the first crossing of the gap in the HRD? This effect is subtle and we have to proceed step by step to understand what causes it. To do so, we study two models of $15\,M_{\sun}$ taken from \citet{Meynet2013a}. These models have the same metallicity as in the present paper but a slightly higher initial rotation velocity, and they differ by the choice of the shear diffusion coefficient (see Table~\ref{Table_HRD_HeBurning}): one with the same prescription as in the present work \citep[hereafter M97]{Maeder1997a}, and one with the prescription by \citet[][hereafter TZ97]{Talon1997a}. First let's note that the sizes of the convective core at the end of the MS phase are very similar. This is not surprising because the size of the convective core is affected mainly through the $D_\text{eff}$ coefficient, which accounts for the concomitant effects of both the meridional currents and the strong horizontal turbulence \citep{Chaboyer1992a}, and which is dominant near the core edge. This coefficient is the same in both models \citep{Zahn1992a}, thus the convective core sizes are very similar.

\begin{figure*}
\centering
\includegraphics[width=.45\textwidth]{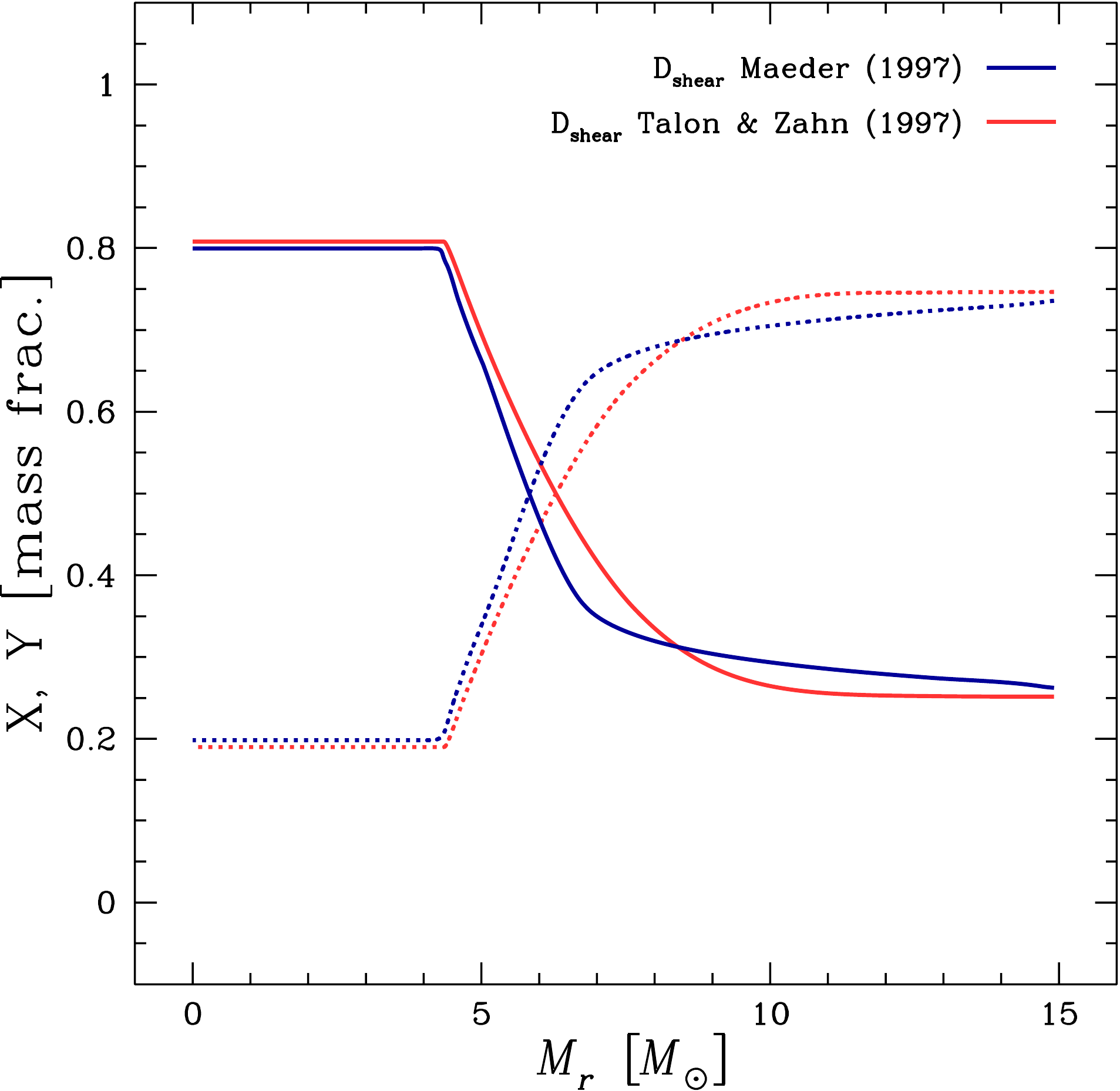}\hspace{.3cm}\includegraphics[width=.45\textwidth]{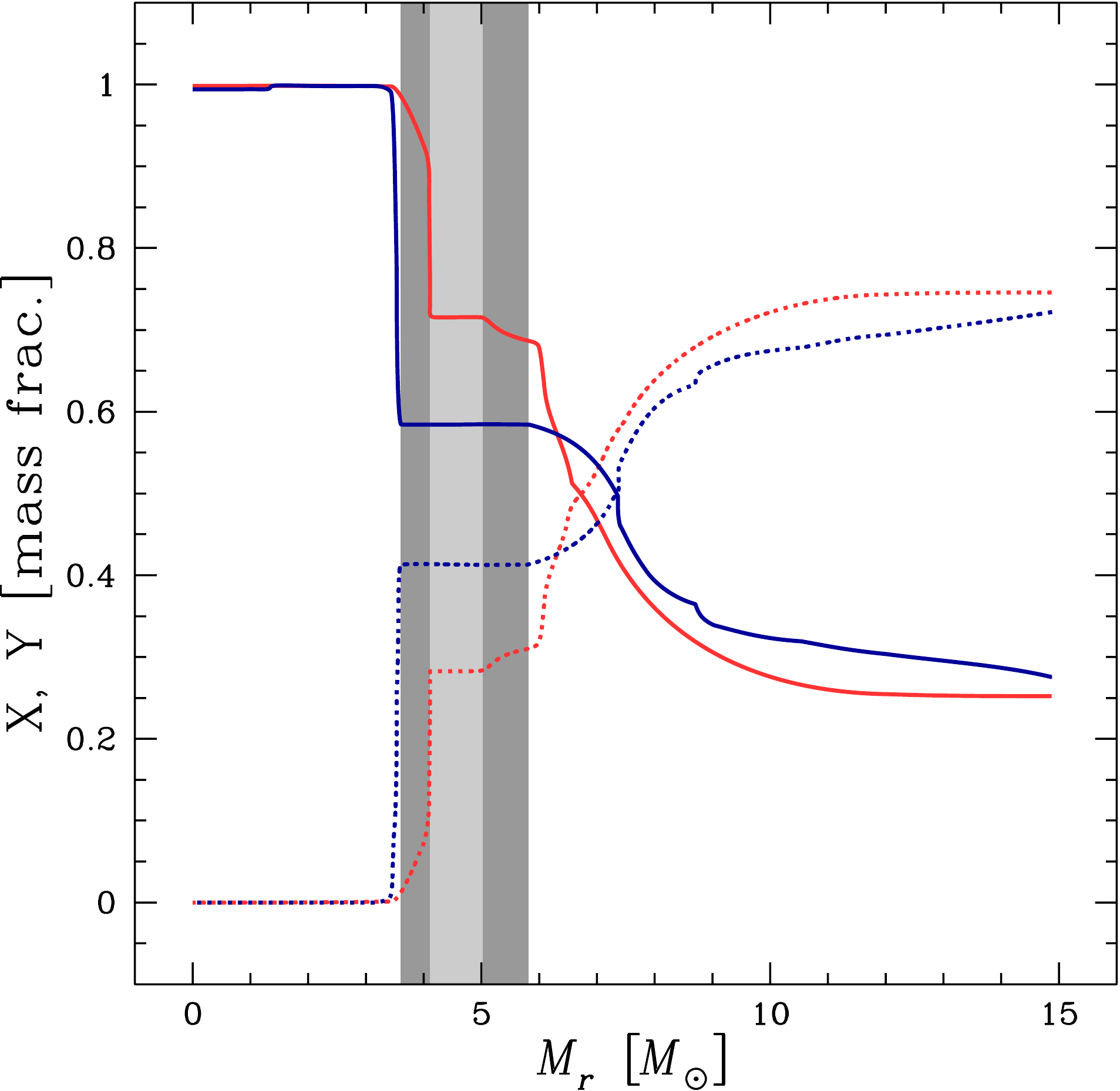}
\caption{Profiles of hydrogen (in mass fraction, dotted lines) and of helium (solid lines) as a function of the Lagrangian mass coordinate inside a $15\,M_{\sun}$ model at $Z=0.002$ with $V_\text{ini}/V_\text{crit}=0.50$ for different prescriptions for the $D_\text{shear}$ (blue: \citetalias{Maeder1997a}; red: \citetalias{Talon1997a}. {\it Left:} profiles in the last third of the MS phase. {\it Right:} just after the MS phase, when  $\log\left(T_\text{eff}\right) =4.3$. The shaded areas represent the extension of the convective zone associated to the H-burning shell (dark grey: \citetalias{Maeder1997a} model; light grey: \citetalias{Talon1997a} model).}
\label{Fig_hy1}
\end{figure*}

Further out from the border of the convective core, the $\mu$-gradients become shallower and the $D_\text{shear}$ becomes the dominant coefficient. The value of the \citetalias{Maeder1997a} $D_\text{shear}$ is higher than the \citetalias{Talon1997a} one. Figure~\ref{Fig_hy1} shows the distributions of hydrogen and helium inside the models. On the left panel, corresponding to the last third of the MS, the blue curves \citepalias{Maeder1997a} are steeper, although they result from the highest values of $D_\text{shear }$. This might appear surprising at first sight: in both models, helium diffuses from the core outside in the radiative envelope in the same way (same $D_\text{eff}$). But in the model with the highest  $D_\text{shear}$, helium is brought further away in a more efficient way, and is mixed in a larger portion of the radiative envelope. The net result is a steeper abundance gradients at the border of the core and a smoother one further out inside the star.

Just after the MS phase and before the ignition of helium in the core (right panel of Fig.~\ref{Fig_hy1}), the chemical configuration is very different in the two models. In the \citetalias{Maeder1997a} model, the convective H-burning shell is wide (dark grey hatched zone), while in the \citetalias{Talon1997a} model, the convective zone is much narrower (light grey hatched zone). The difference in the width of the convective zones comes from two factors:
\begin{enumerate}
\item in the \citetalias{Talon1997a} model, the gradient of helium at the border of the core is shallower, therefore the H-burning shell will rapidly migrate outward and thus will decrease in strength, and the mass of the core will increase more rapidly;
\item the larger helium abundance decrease slightly the opacity.
\end{enumerate}
Both factors tend to give a prominent role to the core for compensating the loss of energy by the surface. The core has to contract more rapidly in this model than in the \citetalias{Maeder1997a} one, and this will produce an expansion of the envelope by mirror effect. In contrast, the H-burning shell of the \citetalias{Maeder1997a} model stays in a deeper region of the star (in mass) where hydrogen remains abundant. This triggers a stronger energy generation which activates a large convective H-burning shell. This convective shell prevents the He-core to increase in mass as fast as in the \citetalias{Talon1997a} model. This results in less energy being released by its contraction and the star keeping a blue position in the HR diagram.

The red to blue evolution is a feature which is very sensitive to many physical ingredients of the models. The fact that it depends on subtle changes in the efficiency of mixing in different regions of the star is just one illustration of this. The bottom line here is that a stronger mixing, not directly above the core, but in a region slightly shifted above the border of the core tends to favour a blue position. Such a stronger mixing makes the gradient of chemical composition steeper at the border of the core and produces a H-burning shell in a deeper position (in mass) in the star, thus with a stronger energy generation and a larger convective shell associated to it.

In the case of the non-rotating models, the difference between the $Z=0.002$ models (which remain in the blue part of the HR diagram during the whole core He-burning phase) and the ones at $Z=0.014$ (which in contrast spend most of their core He-burning phase in the red part of the HR diagram), the explanation is different. The difference comes on one side from the lower opacity and on the other side from the effect of a change in $Z$ on the nuclear energy generation. A lower opacity means that any excess of energy produced by the contracting core will escape more easily. Typically the luminosity  of the $15\,M_{\sun}$ at $Z=0.002$ at $\log\left(T_\text{eff}\right)=4.3$ is $\sim57 900\,L_{\sun}$, \textit{i.e.} $20\%$ larger than the luminosity at the same corresponding evolutionary stage of the $15\,M_{\sun}$ at $Z=0.014$. This is likely the main effect here.

At still lower $Z$, the impact on the nuclear energy generation may be a key effect. The lower abundance of CNO elements forces the H burning to occur at higher temperatures. This means that the core does not need to contract as much as at solar metallicity to reach temperatures sufficient for helium burning at the end of the core hydrogen burning phase. Less energy is released by this contraction and thus less energy is available for inflating the envelope. This is why for instance, Pop III stars spend the whole core He-burning phase in the blue part of the HRD \citep{Ekstrom2008a}. A lower CNO content also implies a deeper position of the H-burning shell and this also favours a blue position in the HRD during a significant part of the core He-burning phase.

\subsection{Comparison with STAREVOL results}

\citet{Lagarde2012a} computed a grid of low- and intermediate-mass stellar models from $0.85$ to $6\,M_{\sun}$ from the PMS up to the AGB phase at the SMC metallicity with the stellar evolution code STAREVOL. At solar metallicity, \citet{Lagarde2012a} show that the use of similar physical inputs (convection, opacity, nuclear reaction rates, mass loss, equation of state, initial abundance) leads to a difference in the main-sequence lifetime of a $4\,M_{\sun}$ non-rotating model of $4\%$ between the two codes. This difference is mainly due to a larger nuclear network in STAREVOL allowing to follow unstable nuclei which produce a smaller core and a lower luminosity. While this effect remains at the SMC metallicity, the difference becomes slightly lower due to the lower initial abundances, so the main-sequence lifetime between two non-rotating models differs only by $2\%$. 

In rotating models, both codes follow the advection of angular momentum by meridional circulation and diffusion by the shear turbulence. However the exact prescription for the horizontal turbulence differ \citep[see][]{Lagarde2012a}, resulting in a lower transport of chemical elements above the convective core in the STAREVOL models. Besides as the STAREVOL models have an initial velocity $25\%$ lower than the present models, the MS lifetime becomes $12\%$ smaller in the rotating $4\,M_{\sun}$ models.

\section{Discussion and conclusion \label{SecDiscu}}

We have studied for the first time at $Z=0.002$, the impact of rotation on the evolution of single stars covering the mass domain from $0.8$ to $120\,M_{\sun}$ in a homogeneous way. The fact that the physics in the present models is the same as in \citetalias{Ekstrom2012a}, where models with an initial metallicity Z=0.014 are examined in detail, allows us to study the impact of a change in the metallicity and in the rotation over an extended mass range. The main results are the following.

\textit{Effect of a change in $Z$ on the non-rotating models:} As found in previous works, we obtain that tracks at $Z=0.002$ are shifted to higher effective temperature and luminosities with respect to those at $Z=0.014$. These shifts are due to the effects of a change in the metallicity on the opacities and on the nuclear energy generation rates. In summary, when the metallicity goes from $Z=0.014$ down to $0.002$:
\begin{itemize}
  \item the widening of the MS occurring in the upper part of the HRD is shifted to a higher luminosity;
  \item the mass range of models presenting a blue loop is extended;
  \item the upper luminosity of RSGs and of the red giant tips remains at the same value;
  \item the mass domain in which the models go through a RSG stage before becoming a WR star is reduced;
  \item for models with $M_\text{ini}\lesssim15\,M_{\sun}$, the MS lifetimes are shorter;
  \item for models with $M_\text{ini}\gtrsim20\, M_{\sun}$, the final masses are $2-3$ times higher.
\end{itemize}

\textit{Effect of rotation ($V_\text{ini}/V_\text{crit}=0.4$) at $Z=0.002$:} With respect to non-rotating models at $Z=0.002$, rotation induces the following characteristics:
\begin{itemize}
  \item tracks no longer show any widening of the MS band in the upper part of the HRD: on the contrary, the MS band becomes narrower;
  \item the mass range of models presenting a blue loop is reduced;
  \item the upper luminosity of RSGs and of the red giant tips remains at the same value
  \item the mass domain in which the models go through a RSG stage before becoming a WR star is extended;
  \item for models with $M_\text{ini}\gtrsim1.7\,M_{\sun}$, the MS lifetimes are increased  by about $20-25\%$
  \item the final masses obtained with rotation show little difference with respect to the values obtained from the models with no rotation, except in the mass range above $50\,M_{\sun}$, where the final masses obtained from the rotating models are higher than those obtained from the non-rotating ones;
  \item the N enrichment at the end of the MS is higher than a factor of three at the surface of the models with $M_\text{ini} > 2.5\,M_{\sun}$, while non-rotating models show no surface enrichments during the whole MS phase for the whole mass range from $0.8$ up to $120\,M_{\sun}$.
\end{itemize}

\textit{Effect of change in $Z$ on the rotating models:} Rotational mixing is more efficient at low $Z$, this is true for massive stars as well as for intermediate- and low-mass stars. The enrichment factors, when the metallicity goes from $Z=0.014$ down to $0.002$, increases when the initial mass decreases, all other characteristics as the initial rotation or the evolutionary stage being taken equal.

The present paper together with \citetalias{Ekstrom2012a} is part of a database of models for $Z=0.002$ and $0.014$. This dataset will be complemented in a near future with the following metallicities: $Z=0.006$ (LMC), $Z=0.00028$ (IZw18), and $Z=0.020$ (Galactic centre). All the data as well as the corresponding isochrones are made available through the web\footnote{\url{http://obswww.unige.ch/Recherche/evol/-Database-} or through the CDS database at  \url{http://vizier.u-strasbg.fr/viz-bin/VizieR-2}.}.

\begin{acknowledgements}
The authors express their gratitude to Dr J. W. Ferguson, who has computed on request the molecular opacities for the peculiar mixture used in this paper. RH acknowledges support from the World Premier International Research Center Initiative (WPI Initiative), MEXT, Japan. CG and RH acknowledge support from the European Research Council under the European Union's Seventh Framework Programme (FP/2007-2013) / ERC Grant Agreement n. 306901. NY acknowledges support from Fundamental Research Grant, Ministry of Higher Education of Malaysia: FP003-2013A.
\end{acknowledgements}

\bibliographystyle{aa}
\bibliography{MyBiblio}

\begin{landscape}
\begin{table}
\caption{Properties of the $Z=0.002$ stellar models at the end of the H-, He-, and C-burning phases.}
\centering
\scalebox{0.75}{\begin{tabular}{rrr|rrrrrr|rrrrrr|rrrrrr}
\hline\hline
\multicolumn{3}{c|}{} & \multicolumn{6}{c|}{End of H-burning} & \multicolumn{6}{c|}{End of He-burning} & \multicolumn{6}{c}{End of C-burning}\\
 $M_\text{ini}$ &  $V_\text{ini}$ & $\bar{V}_\text{MS}$ & $\tau_\text{H}$ & $M$ & $V_\text{eq}$ & $Y_\text{surf}$ & $\text{N}/\text{C}$ & $\text{N}/\text{O}$ & $\tau_\text{He}$ & $M$ & $V_\text{eq}$ & $Y_\text{surf}$ & $\text{N}/\text{C}$ & $\text{N}/\text{O}$ & $\tau_\text{C}$ & $M$ & $V_\text{eq}$ & $Y_\text{surf}$ & $\text{N}/\text{C}$ & $\text{N}/\text{O}$ \\
 $M_{\sun}$ & \multicolumn{2}{c|}{km s$^{-1}$} & Myr & $M_{\sun}$ & km s$^{-1}$ & \multicolumn{3}{c|}{mass fract.} & Myr & $M_{\sun}$ & km s$^{-1}$ & \multicolumn{3}{c|}{mass fract.} & kyr & $M_{\sun}$ & km s$^{-1}$ & \multicolumn{3}{c}{mass fract.}\\
\hline
 $120.00$ & $  0$ & $  0$ & $    2.653$ & $111.76$ & $  0$ & $0.2509$ & $  0.2886$ & $  0.1152$ & $    0.274$ & $ 64.29$ & $  0$ & $0.7888$ & $ 70.7761$ & $101.7192$ & $	0.002$ & $ 56.53$ & $  0$ & $0.9971$ & $ 27.8456$ & $133.9978$ \\
 $120.00$ & $438$ & $322$ & $    3.177$ & $105.61$ & $ 30$ & $0.7462$ & $ 31.9655$ & $ 16.3902$ & $    0.273$ & $ 86.10$ & $  1$ & $0.9247$ & $109.6971$ & $ 87.1787$ & $	0.001$ & $ 85.58$ & $118$ & $0.9523$ & $108.2958$ & $ 92.4621$ \\
 $ 85.00$ & $  0$ & $  0$ & $    3.024$ & $ 80.91$ & $  0$ & $0.2509$ & $  0.2885$ & $  0.1152$ & $    0.305$ & $ 49.00$ & $  0$ & $0.7243$ & $ 72.3763$ & $100.0419$ & $	0.005$ & $ 44.76$ & $  0$ & $0.7340$ & $ 71.4913$ & $101.5953$ \\
 $ 85.00$ & $435$ & $319$ & $    3.594$ & $ 77.24$ & $ 26$ & $0.5375$ & $  9.7941$ & $  3.8372$ & $    0.305$ & $ 51.36$ & $  0$ & $0.8264$ & $132.2048$ & $ 88.9998$ & $	0.002$ & $ 51.13$ & $  0$ & $0.8266$ & $128.9329$ & $ 87.7203$ \\
 $ 60.00$ & $  0$ & $  0$ & $    3.555$ & $ 57.79$ & $  0$ & $0.2509$ & $  0.2885$ & $  0.1152$ & $    0.348$ & $ 38.06$ & $  0$ & $0.6460$ & $ 75.3146$ & $ 98.7258$ & $	0.014$ & $ 35.89$ & $  0$ & $0.6852$ & $ 73.7206$ & $100.2705$ \\
 $ 60.00$ & $400$ & $317$ & $    4.197$ & $ 56.34$ & $ 85$ & $0.4016$ & $  4.7404$ & $  1.6482$ & $    0.350$ & $ 39.45$ & $  0$ & $0.7435$ & $ 64.0025$ & $ 33.7453$ & $	0.006$ & $ 39.12$ & $  0$ & $0.7535$ & $ 70.0064$ & $ 50.9141$ \\
 $ 40.00$ & $  0$ & $  0$ & $    4.507$ & $ 39.10$ & $  0$ & $0.2509$ & $  0.2885$ & $  0.1152$ & $    0.430$ & $ 29.55$ & $  0$ & $0.4943$ & $ 15.4842$ & $  3.4255$ & $	0.069$ & $ 28.35$ & $  0$ & $0.5287$ & $ 40.7668$ & $  5.2130$ \\
 $ 40.00$ & $358$ & $300$ & $    5.253$ & $ 38.67$ & $125$ & $0.3141$ & $  3.2180$ & $  0.9157$ & $    0.448$ & $ 29.16$ & $  0$ & $0.5721$ & $ 20.0054$ & $  4.9839$ & $	0.045$ & $ 27.44$ & $  0$ & $0.6373$ & $ 39.4232$ & $ 12.9620$ \\
 $ 32.00$ & $  0$ & $  0$ & $    5.294$ & $ 31.51$ & $  0$ & $0.2509$ & $  0.2885$ & $  0.1152$ & $    0.520$ & $ 23.91$ & $  0$ & $0.3244$ & $  2.9492$ & $  0.7744$ & $	0.207$ & $ 22.96$ & $  0$ & $0.5435$ & $  1.3441$ & $  5.6032$ \\
 $ 32.00$ & $338$ & $282$ & $    6.354$ & $ 31.24$ & $177$ & $0.3098$ & $  4.0719$ & $  0.9273$ & $    0.543$ & $ 24.68$ & $  7$ & $0.5783$ & $ 28.1622$ & $  5.4938$ & $	0.107$ & $ 24.07$ & $  1$ & $0.5971$ & $ 33.5561$ & $  7.0614$ \\
 $ 25.00$ & $  0$ & $  0$ & $    6.449$ & $ 24.78$ & $  0$ & $0.2509$ & $  0.2885$ & $  0.1152$ & $    0.655$ & $ 24.40$ & $  0$ & $0.2509$ & $  0.2885$ & $  0.1152$ & $	0.370$ & $ 24.07$ & $  0$ & $0.2509$ & $  0.2886$ & $  0.1152$ \\
 $ 25.00$ & $319$ & $262$ & $    7.663$ & $ 24.67$ & $132$ & $0.2923$ & $  4.6546$ & $  0.8616$ & $    0.694$ & $ 22.77$ & $  1$ & $0.3161$ & $  6.3721$ & $  1.0360$ & $	0.256$ & $ 21.83$ & $  0$ & $0.3222$ & $  6.8124$ & $  1.0774$ \\
 $ 20.00$ & $  0$ & $  0$ & $    7.888$ & $ 19.91$ & $  0$ & $0.2509$ & $  0.2885$ & $  0.1152$ & $    0.871$ & $ 19.70$ & $  0$ & $0.2509$ & $  0.2885$ & $  0.1152$ & $	0.845$ & $ 19.34$ & $  0$ & $0.3514$ & $  3.4815$ & $  0.9683$ \\
 $ 20.00$ & $305$ & $248$ & $    9.238$ & $ 19.87$ & $194$ & $0.2772$ & $  4.7364$ & $  0.7916$ & $    0.945$ & $ 19.25$ & $  2$ & $0.2851$ & $  5.6625$ & $  0.8651$ & $	0.630$ & $ 18.66$ & $  1$ & $0.3019$ & $  7.5887$ & $  0.9985$ \\
 $ 15.00$ & $  0$ & $  0$ & $   10.996$ & $ 14.92$ & $  0$ & $0.2509$ & $  0.2885$ & $  0.1152$ & $    1.287$ & $ 14.78$ & $  0$ & $0.2510$ & $  0.8404$ & $  0.2386$ & $	3.324$ & $ 14.68$ & $  0$ & $0.3106$ & $  2.6928$ & $  0.6531$ \\
 $ 15.00$ & $303$ & $227$ & $   13.241$ & $ 14.88$ & $225$ & $0.2825$ & $  6.5806$ & $  0.8607$ & $    1.349$ & $ 14.77$ & $  4$ & $0.2844$ & $  6.8636$ & $  0.8768$ & $	2.834$ & $ 14.67$ & $  1$ & $0.3878$ & $ 19.9244$ & $  1.8530$ \\
 $ 12.00$ & $  0$ & $  0$ & $   15.100$ & $ 11.97$ & $  0$ & $0.2509$ & $  0.2885$ & $  0.1152$ & $    1.902$ & $ 11.91$ & $  0$ & $0.2509$ & $  0.9591$ & $  0.2576$ & $	4.541$ & $ 11.84$ & $  0$ & $0.2822$ & $  2.3020$ & $  0.5332$ \\
 $ 12.00$ & $271$ & $216$ & $   18.130$ & $ 11.95$ & $219$ & $0.2775$ & $  6.7399$ & $  0.8378$ & $    1.911$ & $ 11.88$ & $  2$ & $0.2992$ & $ 10.7353$ & $  1.0064$ & $	5.350$ & $ 11.78$ & $  1$ & $0.3520$ & $ 17.4150$ & $  1.3434$ \\
 $  9.00$ & $  0$ & $  0$ & $   25.471$ & $  8.99$ & $  0$ & $0.2509$ & $  0.2885$ & $  0.1152$ & $    2.999$ & $  8.89$ & $  0$ & $0.2510$ & $  1.2200$ & $  0.2935$ &  \\
 $  9.00$ & $255$ & $203$ & $   30.205$ & $  8.99$ & $208$ & $0.2718$ & $  6.1934$ & $  0.7959$ & $    3.249$ & $  8.98$ & $ 11$ & $0.2727$ & $  6.3887$ & $  0.8058$ & $	3.683$ & $  8.92$ & $  1$ & $0.3114$ & $ 15.1826$ & $  1.1064$ \\
 \cline{16-21}
 $  7.00$ & $  0$ & $  0$ & $   39.945$ & $  7.00$ & $  0$ & $0.2509$ & $  0.2885$ & $  0.1152$ & $    5.570$ & $  6.94$ & $  0$ & $0.2511$ & $  1.0737$ & $  0.2745$ &  \\
 $  7.00$ & $264$ & $194$ & $   46.954$ & $  7.00$ & $198$ & $0.2663$ & $  4.5964$ & $  0.7037$ & $    5.360$ & $  6.95$ & $ 29$ & $0.2848$ & $  8.3414$ & $  0.8934$ &  \\
 $  5.00$ & $  0$ & $  0$ & $   77.356$ & $  5.00$ & $  0$ & $0.2509$ & $  0.2885$ & $  0.1152$ & $   13.652$ & $  4.96$ & $  0$ & $0.2513$ & $  1.0048$ & $  0.2649$ &  \\
 $  5.00$ & $228$ & $182$ & $   91.943$ & $  5.00$ & $185$ & $0.2612$ & $  2.9866$ & $  0.5854$ & $   12.916$ & $  4.96$ & $  4$ & $0.2814$ & $  6.4353$ & $  0.8321$ &  \\
 $  4.00$ & $  0$ & $  0$ & $  124.375$ & $  4.00$ & $  0$ & $0.2509$ & $  0.2885$ & $  0.1152$ & $   26.143$ & $  3.97$ & $  0$ & $0.2519$ & $  1.1545$ & $  0.2856$ &  \\
 $  4.00$ & $230$ & $174$ & $  148.904$ & $  4.00$ & $176$ & $0.2585$ & $  2.2090$ & $  0.4910$ & $   23.675$ & $  3.96$ & $  4$ & $0.2824$ & $  5.8138$ & $  0.7906$ &  \\
 $  3.00$ & $  0$ & $  0$ & $  240.308$ & $  3.00$ & $  0$ & $0.2509$ & $  0.2885$ & $  0.1152$ & $   63.990$ & $  2.97$ & $  0$ & $0.2549$ & $  1.4165$ & $  0.3226$ &  \\
 $  3.00$ & $207$ & $166$ & $  290.180$ & $  3.00$ & $162$ & $0.2566$ & $  1.5848$ & $  0.3969$ & $   55.093$ & $  2.96$ & $  5$ & $0.2898$ & $  5.8778$ & $  0.7796$ &  \\
 $  2.50$ & $  0$ & $  0$ & $  374.191$ & $  2.50$ & $  0$ & $0.2509$ & $  0.2885$ & $  0.1152$ & $  114.955$ & $  2.48$ & $  0$ & $0.2648$ & $  1.7337$ & $  0.3915$ &  \\
 $  2.50$ & $198$ & $160$ & $  457.047$ & $  2.50$ & $156$ & $0.2562$ & $  1.3327$ & $  0.3502$ & $   97.943$ & $  2.47$ & $  5$ & $0.2988$ & $  5.9972$ & $  0.7729$ &  \\
 $  2.00$ & $  0$ & $  0$ & $  667.029$ & $  2.00$ & $  0$ & $0.2509$ & $  0.2885$ & $  0.1152$ & & & & & & & \\
 $  2.00$ & $213$ & $158$ & $  833.680$ & $  2.00$ & $158$ & $0.2570$ & $  1.1280$ & $  0.3015$ & $  189.820$ & $  1.97$ & $  5$ & $0.3050$ & $  5.6331$ & $  0.6926$ &  \\
\cline{10-15}
 $  1.70$ & $  0$ & $  0$ & $ 1053.592$ & $  1.70$ & $  0$ & $0.2509$ & $  0.2885$ & $  0.1152$ &  \\
 $  1.70$ & $201$ & $154$ & $ 1322.276$ & $  1.70$ & $157$ & $0.2576$ & $  0.9529$ & $  0.2639$ &  \\
 $  1.50$ & $  0$ & $  0$ & $ 1445.884$ & $  1.50$ & $  0$ & $0.2509$ & $  0.2885$ & $  0.1152$ &  \\
 $  1.50$ & $148$ & $139$ & $ 1592.012$ & $  1.50$ & $146$ & $0.2511$ & $  0.2934$ & $  0.1164$ &  \\
 $  1.35$ & $  0$ & $  0$ & $ 1985.595$ & $  1.35$ & $  0$ & $0.2509$ & $  0.2885$ & $  0.1152$ &  \\
 $  1.35$ & $ 100$ & $ 94$ & $ 2083.188$ & $  1.35$ & $101$ & $0.2509$ & $  0.2886$ & $  0.1152$ &  \\
 $  1.25$ & $  0$ & $  0$ & $ 2600.008$ & $  1.25$ & $  0$ & $0.2509$ & $  0.2885$ & $  0.1152$ &  \\
 $  1.25$ & $100$ & $  6$ & $ 2628.941$ & $  1.25$ & $  5$ & $0.2562$ & $  0.5579$ & $  0.1821$ &  \\
 $  1.10$ & $  0$ & $  0$ & $ 3796.321$ & $  1.10$ & $  0$ & $0.2509$ & $  0.2885$ & $  0.1152$ &  \\
 $  1.10$ & $ 50$ & $  4$ & $ 3822.563$ & $  1.10$ & $  4$ & $0.2535$ & $  0.3153$ & $  0.1225$ &  \\
 $  1.00$ & $  0$ & $  0$ & $ 5491.634$ & $  1.00$ & $  0$ & $0.2509$ & $  0.2885$ & $  0.1152$ &  \\
 $  1.00$ & $ 50$ & $  3$ & $ 5564.119$ & $  1.00$ & $  3$ & $0.2543$ & $  0.3121$ & $  0.1216$ &  \\
 $  0.90$ & $  0$ & $  0$ & $ 7315.457$ & $  0.90$ & $  0$ & $0.0900$ & $  0.3002$ & $  0.1116$ &  \\
 $  0.90$ & $  50$ & $  2$ & $ 7470.584$ & $  0.90$ & $  2$ & $0.2391$ & $  0.2980$ & $  0.1171$ &  \\
 $  0.80$ & $  0$ & $  0$ & $11574.902$ & $  0.80$ & $  0$ & $0.1514$ & $  0.2943$ & $  0.1134$ &  \\
 $  0.80$ & $  50$ & $  1$ & $11843.692$ & $  0.80$ & $  1$ & $0.2340$ & $  0.2941$ & $  0.1159$ &  \\
\cline{1-9}
\end{tabular}}
\tablefoot{The non-rotating $2\,M_{\sun}$ undergoes He-flash and the non-rotating $9\,M_{\sun}$ does not succeed to go through the core C-burning phase.}
\label{TabListModels}
\end{table}
\end{landscape}

\end{document}